\documentclass[12pt,useAMS,usenatbib]{emulateapj}

\usepackage{natbib}

\usepackage{graphicx}
\usepackage{amsmath}
\usepackage{color}

\newcommand{\Msun}{M_{\odot}}

\newcommand{\Mtot}{M_{\rm tot}}
\newcommand{\fesc}{f_{\rm esc}}

\newcommand{\art}{\rm ART^{2}}
\newcommand{\A}{\rm \AA}
\newcommand{\La}{L_{\rm{Ly\alpha}}}

\newcommand{\EW}{\rm EW}

\newcommand{\Mpc}{\rm {Mpc}}

\newcommand{\MBH}{{\rm M_{\rm{BH}}}}

\newcommand{\Msunyr}{\rm {M_{\odot}\; yr^{-1}}}

\newcommand{\ergs}{{\rm erg~s^{-1}}}
\newcommand{\fa}{f_{\rm LAE}}

\newcommand{\lya}{\ifmmode {\rm Ly}\alpha \else Ly$\alpha$\fi}
\def\msunyr{\ifmmode M_{\odot} {\rm yr}^{-1} \else M$_{\odot}$ yr$^{-1}$\fi}

\begin{document}                          
%
%

\title{WERE PROGENITORS OF LOCAL L* GALAXIES $\lya$ EMITTERS AT HIGH REDSHIFT?}    

%
%
\author
{
Hidenobu Yajima\altaffilmark{1,2},
Yuexing Li\altaffilmark{1,2},
Qirong Zhu\altaffilmark{1,2}, 
Tom Abel\altaffilmark{3},
Caryl Gronwall\altaffilmark{1,2},
Robin Ciardullo\altaffilmark{1,2}  
}

\affil{$^{1}$Department of Astronomy \& Astrophysics, The Pennsylvania State University, 
525 Davey Lab, University Park, PA 16802, USA}
\affil{$^{2}$Institute for Gravitation and the Cosmos, The Pennsylvania State University, University Park, PA 16802}
\affil{$^{3}$Kavli Institute for Particle Astrophysics and Cosmology, SLAC National Accelerator Laboratory, Stanford University, \\
2575 Sand Hill Road, Menlo Park, CA 94025, USA}

\email{yuh19@psu.edu} 

%
%

\begin{abstract}
The $\lya$ emission has been observed from galaxies over a redshift span $z \sim 0 - 8.6$. However, the evolution of high-redshift $\lya$ emitters (LAEs), and the link between these populations and local galaxies, remain poorly understood. Here, we investigate the $\lya$ properties of progenitors of a local $L^{*}$ galaxy by combining cosmological hydrodynamic simulations with three-dimensional radiative transfer calculations using the new $\art$ code. We find that the main progenitor (the most massive one) of a Milky Way-like galaxy has a number of $\lya$ properties close to those of observed LAEs at $z \sim 2 - 6$, but most of the fainter ones appear to fall below the detection limits of current surveys. The $\lya$ photon escape fraction depends sensitively on a number of physical properties of the galaxy, such as mass, star formation rate, and metallicity, as well as galaxy morphology and orientation.  Moreover, we find that high-redshift LAEs show blue-shifted $\lya$ line profiles characteristic of gas inflow, and that the $\lya$ emission by excitation cooling increases with redshift, and becomes dominant at $z \gtrsim 6$. Our results suggest that some observed LAEs at $z \sim 2-6$ with luminosity of $\La \sim 10^{42-43}~\ergs$ may be similar to the main progenitor of the Milky Way at high redshift, and that they may evolve into present-day $L^{*}$ galaxies.

\end{abstract}


%
%

\section{INTRODUCTION}

The $\lya$ emission from young galaxies can be a powerful probe of the early universe \citep{Partridge67, Charlot93}. Recent narrow-band deep imaging surveys using large-aperture telescopes have detected a large number of $\lya$ emitting galaxies, or  $\lya$ emitters (LAEs), at redshifts $z \gtrsim 3$ \citep[e.g.,][]{Hu96, Cowie98, Steidel00, Malhotra04, Taniguchi05, Kashikawa06, Shimasaku06, Iye06, Hu2006, Gronwall07, Ouchi08, Hu2010, Ouchi10, Lehnert10}. By combining $\lya$ emission with broad-band continuum, multi-wavelength observations are beginning to address the physical properties of these high-redshift LAEs \citep[e.g.,][]{Gawiser06, Gronwall07, Lai07, Nilsson07, Pirzkal07, Lai08, Ouchi08, Pentericci09, Ono10A, Ono10B, Hayes10, Finkelstein11, Nilsson11, Acquaviva11}. It has been suggested that these objects are mostly compact, young galaxies with low metallicity. In addition, \citet{Ouchi08} studied the evolution of equivalent widths (EWs) and the characteristic $L^{*}_{\lya}$ with redshift from $z \sim 3$ to $z\sim 6$, and found that the mean EW increased with redshift, while the $L^{*}_{\lya}$ did not change significantly. More recently, \citet{Ciardullo11} studied the evolution of luminosity function (LF) and EW from $z=2.1$ to $3.1$, and found that $L_{*}$ increases from $10^{42.3}~\rm \ergs$ at $z=2.1$ to $10^{43}~\rm \ergs$ at $z=3.1$. \citet{Blanc11} studied the $\lya$ properties of LAEs in the redshift range $z=1.9-3.8$ from the Hobby Eberly Telescope Dark Energy Experiment (HETDEX) Pilot Survey, and showed that the median $\lya$ escape fraction ($\fesc$) was $\sim 29 \%$, and it does not evolve significantly with redshift. On the other hand, \citet{Hayes11} suggested that $\fesc$ monotonically increases between redshift 0 and 6, which implies that high-z galaxies tend to be LAEs.

While high-redshift LAEs have been studied with large samples in the redshift range of $z \sim 2.2-6.6$, there is only a limited number of observations on LAEs at $z \lesssim 1$. Some local star-forming galaxies have been studied by various wavelengths and show a complex structure of $\lya$ and UV continuum \citep{Hayes07, Ostlin09}. \citet{Atek09} showed that the $\fesc$ of local LAEs have a large dispersion, ranging from $\sim 3$ to $100~\%$. In addition, \citet{Deharveng08} studied a sample of 96 local LAEs at $z=0.2-0.35$ from UV space telescope {\it GALEX}, and found that these LAEs have similar EW distribution as those at $z=3.1$. Recently, \citet{Cowie10} have studied $z \sim 0.3$ LAEs with a larger sample, and showed that these LAEs are more compact, and have lower metallicity than UV-continuum selected galaxies at the same redshift. In addition, \citet{Finkelstein09} suggested, from fitting of spectral energy distributions (SEDs), that low-z LAEs are significantly more massive and older galaxies than their high-z counterparts.

One of the important issues in galaxy evolution is how high-redshift LAEs evolve into galaxies in the local universe. \citet{Gawiser07} suggested, from clustering analysis, that most $z=3.1$ LAEs evolve to present-day galaxies of $\lesssim 2.5~ L^{*}$, unlike other populations which typically evolve into more massive galaxies. Moreover, \citet{Guaita10} indicated that LAEs at $z=2.1$ were building blocks of present-day $L^{*}$ galaxies such as the Milky Way (MW). 

However, the link between high-redshift LAEs and local galaxies, and the probability of these LAEs evolving into present-day $L^{*}$ galaxies are not well constrained from observations. In order to address these questions, one may use the Milky Way as a local laboratory. Moreover, since $\lya$ emission has been detected from the most distant galaxies, understanding of the $\lya$ properties of the Milky Way progenitors will provide an important clue to the formation of early galaxies. To date, there are only a limited number of theoretical studies on this important issue \citep[e.g.,][]{Salvadori10, Dayal11b}. Both \citet{Salvadori10} and \citet{Dayal11b} focused on MW progenitors at $z\sim 6$ constructed from semi-analytical merger trees and a cosmological smoothed particle hydrodynamics (SPH) simulation, respectively. They both used the same analytical prescription of $\lya$ emission in which the intrinsic $\lya$ luminosity scales linearly with the star formation rate \citep{Dayal08}. However, because the $\lya$ properties depend sensitively on a number of factors, such as the scattering and propagation of the photons in the inhomogeneous medium, the dust content of the gas, the ionization structure, the UV continuum, and the photon escape fraction. Such a complicated process can only be probed by comprehensive $\lya$ radiative transfer calculations combined with realistic simulation of galaxy formation. As we will show in this work, our detailed $\lya$ modeling on a high-resolution cosmological simulation produce a number of $\lya$ properties such as the luminosity functions at different redshifts in good agreement with observations. Moreover, in order to investigate the evolution of LAEs, we need to study the progenitors of the MW at different redshifts systematically, not just at a specific time.   

In this paper, we investigate the $\lya$ properties of MW progenitors over a wide redshift range of $z \sim 0 - 10$, by combing cosmological SPH simulation of a MW-like galaxy from Zhu et al. (in preparation) with 3D RT calculations using the newly improved $\art$ code by \cite{Yajima11A}. Our hydrodynamic simulation includes important physics of both dark and baryonic matter, and has high resolutions to track the formation history of the MW. Our RT calculations include both $\lya$ resonant scattering and continuum emission, and are done on an adaptive-mesh refinement grid, which covers a large dynamical range and resolves the small-scale structures in high-density region. Interstellar dust is also taken into account to accurately estimate the $\fesc$ of $\lya$ photons and UV continuum, and the EWs.

The paper is organized as follows. We describe our cosmological simulation in \S2, and the RT calculations in \S3. In \S4, we present results of the $\lya$ properties of MW progenitors from redshift 10 to 0, which include the $\lya$ surface brightness, $\lya$ luminosity, $\fesc$, EW and line profile. 
In \S5, we discuss the dependence of $\fesc$ on physical properties, LAE fraction in our galaxy sample, $\lya$ escaping angle and the contribution from excitation cooling to $\lya$ emissivity, and we summarize in \S6. 

%
%


\section{Galaxy model}
The cosmological simulation presented here follows the formation and evolution of a Milky Way-size galaxy and its substructures, as described in detail in Zhu et al. (in preparation). The simulation includes dark matter, gas dynamics, star formation, black hole growth, and feedback processes. The initial condition is originally from the Aquarius Project \citep{Springel08a}, which produced the largest ever particle simulation of a Milky Way-sized dark matter halo. The hydrodynamical initial condition is reconstructed from the original collisionless one by splitting each original particle into a dark matter and gas particle pair, as adopted from \cite{Wadepuhl11}.

 The whole simulation falls in a periodic box of $100~h^{-1} \Mpc$ on each side with a zoom-in region of a size $5\times 5\times 5~h^{-3}\Mpc^{3}$. The spatial resolution is $\sim 250~h^{-1}$ pc in the zoom-in region. The mass resolution of this zoom-in region is $1.8 \times 10^{6}~ h^{-1} \Msun$ for dark matter particles, $3  \times  10^{5}~ h^{-1} \Msun$ for gas, and  $1.5 \times 10^{5}~ h^{-1} \Msun$ for star particles. The cosmological parameters used in the simulation are $\Omega_{m }= 0.25$, $\Omega_{\Lambda} = 0.75$, $\sigma_{8} = 0.9$ and $h=0.73$, consistent with the five-year results of the WMAP \citep{Komatsu09}. The simulation evolves from $z = 127$ to $z = 0$.

The simulation was performed using the parallel, N-body/SPH code GADGET-3, which is an improved version of that described in \cite{Springel01, Springel05e}. For the computation of gravitational forces, the code uses the ``TreePM'' method \citep{Xu95} that combines a ``tree'' algorithm \citep{Barnes86} for short-range forces and a Fourier transform particle-mesh method \citep{Hockney81} for long-range forces. GADGET implements the entropy-conserving formulation of SPH \citep{Springel02} with adaptive particle smoothing, as in \cite{Hernquist89}. 
Radiative cooling and heating processes are calculated assuming collisional ionization equilibrium \citep{Katz96, Dave99}. Star formation is modeled in a multi-phase ISM, with a rate that follows the Schmidt-Kennicutt Law (\citealt{Schmidt59, Kennicutt98}). Feedback from supernovae is captured through a multi-phase model of the ISM by an effective equation of state for star-forming gas \citep{Springel03a}. The UV background model of \cite{Haardt96} is used. 

Black hole growth and feedback are also included in our simulation based on the model of \cite{Springel05d, DiMatteo05}, where black holes are represented by collisionless ``sink'' particles that interact gravitationally with other components and accrete gas from their surroundings. The accretion rate is estimated from the local gas density and sound speed using a spherical Bondi-Hoyle \citep{Bondi52, Bondi44, Hoyle41} model that is limited by the Eddington rate. Feedback from black hole accretion is modeled as thermal energy, $\sim 5\%$ of the radiation, injected into surrounding gas isotropically, as described in \cite{Springel05d} and \cite{DiMatteo05}. This feedback scheme self-regulates the growth of the black hole and has been demonstrated to successfully reproduce many observed properties of local elliptical galaxies (e.g,, \citealt{Springel05a, Hopkins06}) and the most distant quasars at $z \sim 6$ \citep{Li07}. We follow the black hole seeding scheme of \cite{Li07} and \cite{DiMatteo08} in the simulation: a seed black hole of mass ${\MBH} = 10^{5}~ h^{-1} \Msun$ was planted in the gravitational potential minimum of each new halo identified by the friends-of-friends (FOF) group finding algorithms with a total mass greater than $10^{10}~ h^{-1} \Msun$. 

The galaxies in each snapshot above redshift 0 are building blocks of the MW, a present-day $L^{*}$ galaxy. We therefore define them as ``progenitors'' of the MW, and the most massive progenitor at any given timestep as the ``main progenitor''. In this paper, we explore the $\lya$ properties of $\sim 60$ most massive progenitors of each snapshot for 15 snapshots in the redshift range $z=0-10.2$, which gives a total sample of 941 galaxies.

\section{Radiative Transfer}

The RT calculations are done using the 3D Monte Carlo RT code, All-wavelength Radiative Transfer with Adaptive Refinement Tree ($\art$), as recently developed by \cite{Yajima11A}. $\art$ was improved over the original version of \cite{Li08}, and features three essential modules: continuum emission from X-ray to radio, $\lya$ emission from both recombination and collisional excitation, and ionization of neutral hydrogen. The coupling of these three modules, together with an adaptive refinement grid, enables a self-consistent and accurate calculation of the $\lya$ properties, which depend strongly on the UV continuum, ionization structure, and dust content of the object. Moreover, it efficiently produces multi-wavelength properties, such as the spectral energy distribution and images, for direct comparison with multi-band observations. The detailed implementations of the $\art$ code are described in \cite{Li08} and \cite{Yajima11A}. Here we focus on the $\lya$ calculations and briefly outline the process. 

The $\lya$ emission is generated by two major mechanisms: recombination of ionizing photons and collisional excitation of hydrogen gas. In the recombination process, we consider ionization of neutral hydrogen by ionizing radiation from stars, active galactic nucleus (AGN), and UV background (UVB), as well as by collisions by high-temperature gas. The ionized hydrogen atoms then recombine and create $\lya$ photons via the state transition $\rm 2P \rightarrow 1S$. The $\lya$ emissivity from the recombination is
\begin{equation}
\epsilon^{\rm rec}_{\alpha} = f_{\alpha } \alpha_{\rm B} h \nu_{\rm \alpha} n_{\rm e} n_{\rm HII},
\end{equation}
where $\alpha_{\rm B}$ is the case B recombination coefficient, and $f_{\alpha}$ is the average number of $\lya$ photons produced per case B recombination. Here we use $\alpha_{\rm B}$ derived in \citet{Hui97}.
Since the temperature dependence of $f_{\alpha}$ is not strong, $f_{\alpha} = 0.68$ is assumed everywhere \citep{Osterbrock06}. The product $h\nu_{\alpha}$ is the energy of a $\lya$ photon, 10.2 eV.
 
In the process of collisional excitation, high temperature electrons can excite the quantum state of hydrogen gas by the collision. Due to the large Einstein A coefficient, the hydrogen gas can occur de-excitation with the $\lya$ emission. The $\lya$ emissivity by the collisional excitation is estimated by
\begin{equation}
\epsilon^{\rm coll}_{\alpha} = C_{\rm Ly \alpha} n_{\rm e} n_{\rm HI},
\end{equation}
where $C_{\rm Ly \alpha}$ is the collisional excitation coefficient,
$C_{\rm Ly\alpha} = 3.7 \times 10^{-17} {\rm exp}(- h\nu_{\alpha}/kT) T^{-1/2}~\rm ergs\; s^{-1}\; cm^{3}$ \citep{Osterbrock06}.

Once the ionization structure have been determined, we estimate the intrinsic $\lya$ emissivity in each cell by the sum of above $\lya$ emissivity, $\epsilon_{\alpha} = \epsilon^{\rm rec}_{\alpha} + \epsilon^{\rm coll}_{\alpha}$. 

In RT calculations, dust extinction from the ISM is included. The dust content is estimated according to the gas content and metallicity in each cell, which are taken from the hydrodynamic simulation. The dust-to-gas ratio of the MW is used where the metallicity is of Solar abundance, and it is linearly interpolated for other metallicity. We use the stellar population synthesis model of GALAXEV \citep{Bruzual03} to produce intrinsic SEDs of stars for a grid of metallicity and age, and we use a simple, broken power law for the AGN \citep{Li08}.
A \citet{Salpeter55} initial mass function is used in our calculations. 

In this work, we apply $\art$ to the 60 most massive progenitors of each snapshot for 15 snapshots at redshifts spanning $z=0-10.2$. In our post-processing procedure, we first calculate the RT of ionizing photons ($\lambda \le 912 \;\rm \AA$) and estimate the ionization fraction of the ISM. The resulting ionization structure is then used to run the $\lya$ RT to derive the emissivity, followed by the calculation of non-ionizing continuum photons ($\lambda > 912 \;\rm \AA$) in each cell. Our fiducial run is done with $N_{\rm ph} = 10^{5}$ photon packets for each ionizing, $\lya$, and non-ionizing components. Because the spatial resolution of the cosmological simulation is not adequate to resolve the multiple phase of the ISM, we assume a single-phase medium in each density grid. The highest refinement of the grid corresponds to a cell size comparable to the spatial resolution of 250 pc in comoving coordinate of the hydrodynamic simulation.

%
%

\section{Results}

\subsection{The Formation History of The Milky Way}

\begin{figure*}
\begin{center}
\includegraphics[scale=1.0, bb=75 220 531 521, clip=true]{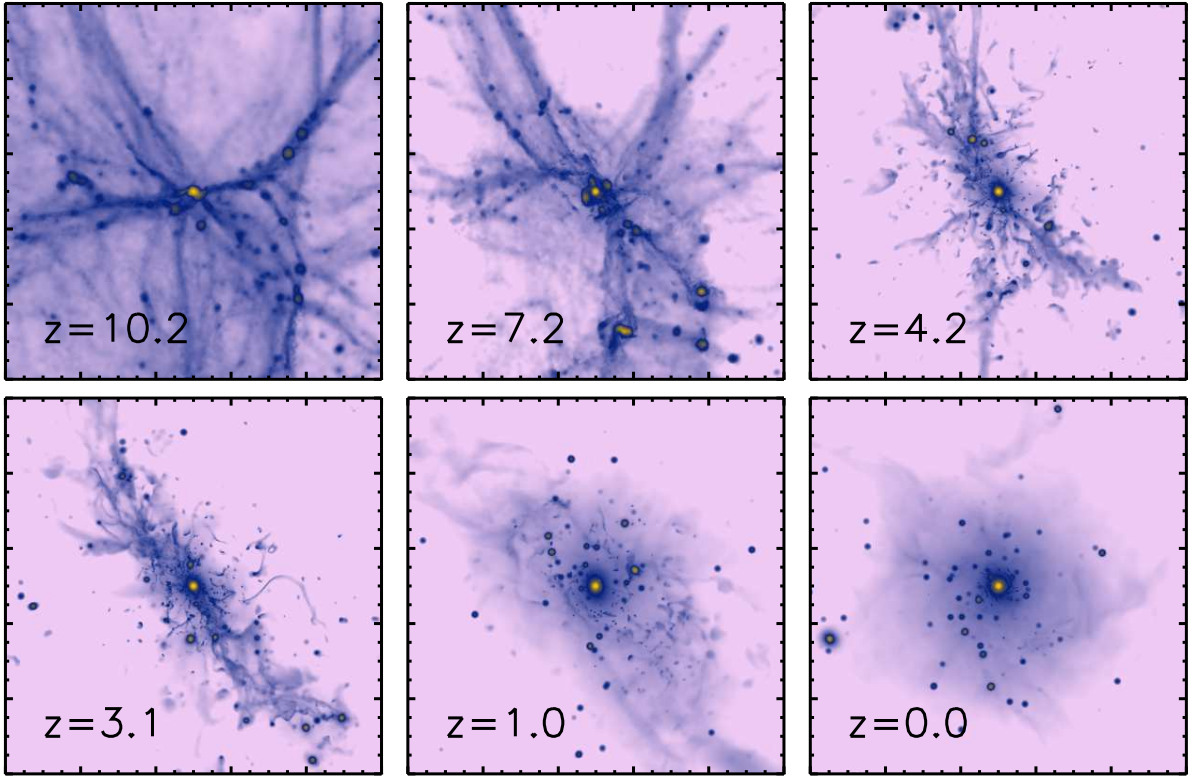}
\caption{Evolution of the MW galaxy from redshift z=10 to z=0. The images are projected density of gas and stars. 
The box size is 1 Mpc in comoving scale. The gas follows the distribution of dark matter and shows filamentary structures. 
Stars and galaxies form in high density regions along the filaments.
}
\label{fig:mw}
\end{center}
\end{figure*}

\begin{figure}
\begin{center}
\includegraphics[scale=0.42]{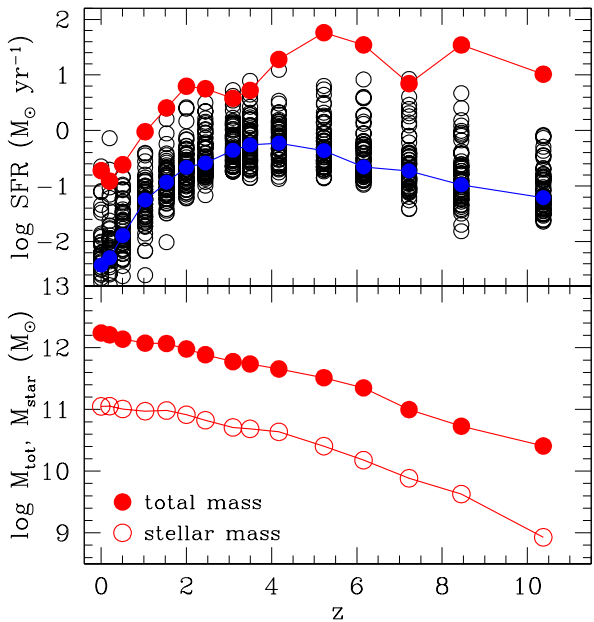}
\caption{The growth history of the MW galaxy. 
Top panel: 
the individual star formation history of the most massive 60 progenitors at each redshift. Red filled circles represent the main progenitor (the most massive one), 
and the blue filled circles indicate the median value of our sample galaxies represented by black open circles.
Bottom panel: 
the accumulated mass of the main progenitor at different redshift. Filled circle represents the total mass, while the open circle represents the stellar mass. 
}
\label{fig:sfr}
\end{center}
\end{figure}

Figure~\ref{fig:mw} shows the evolution of the MW galaxy from redshift z=10 to z=0 from the cosmological simulation. The gas follows the distribution of dark matter in filamentary structures, and stars form in high density regions along the filaments. The most massive galaxy resides in the intersection of the filaments, the highest density peak in the simulated volume where gas concentrates in the deep potential well. The MW galaxy is formed by gas accretion and merger of subhalos, it has the last major merger at redshift $z \sim 2$ (Zhu et al, in preparation). 

Stars start to form at $z \sim 15$ by accretion of cold gas. As shown in Figure~\ref{fig:sfr}, the star formation rate (SFR) of the main progenitor reaches $\sim 10 ~\Msunyr$ at $z \sim 10$, and it peaks at $\sim 58 ~\Msunyr$ at $z = 5.2$, owing to merger of gas-rich protogalaxies. Galaxy interaction induces gravitational torques and shocks, which trigger global starburst. After that, the SFR generally decreases except a boosted bump at $z \sim 2$ when the last major merger takes place. 

Shown also in Figure~\ref{fig:sfr} are the SFRs of the top 60 progenitors at different redshifts. The median and total value of these SFRs gradually increase until it reaches the peak at $z \sim 3.5$, after which it decreases rapidly by over an order of magnitude. This evolution is in broad agreement with the observed cosmic star formation history \citep{Hopkins2006}, although the simulation box is somewhat small to discuss such a statistical property. The star formation at high redshifts ($z \gtrsim 4$) is largely fueled by inflow of cold gas and mergers of gas-rich halos, while the rapid decline of SFR at $z \lesssim 2$ is mainly caused by feedback from stars and AGNs, and the depletion of cold gas.

The main progenitor has a total mass of $\sim 2.6\times 10^{10}\, \Msun$, and a stellar mass of $\sim 8.5\times 10^{8}\, \Msun$ at $z \sim 10$. It evolved into a disk galaxy at $z=0$ with a total mass of $1.6 \times 10^{12}~\Msun$ and stellar mass of $\sim 10^{11}\, \Msun$, as observed in the MW galaxy.

\subsection{$\lya$ Surface Brightness}

\begin{figure*}
\begin{center}
\includegraphics[scale=1.1]{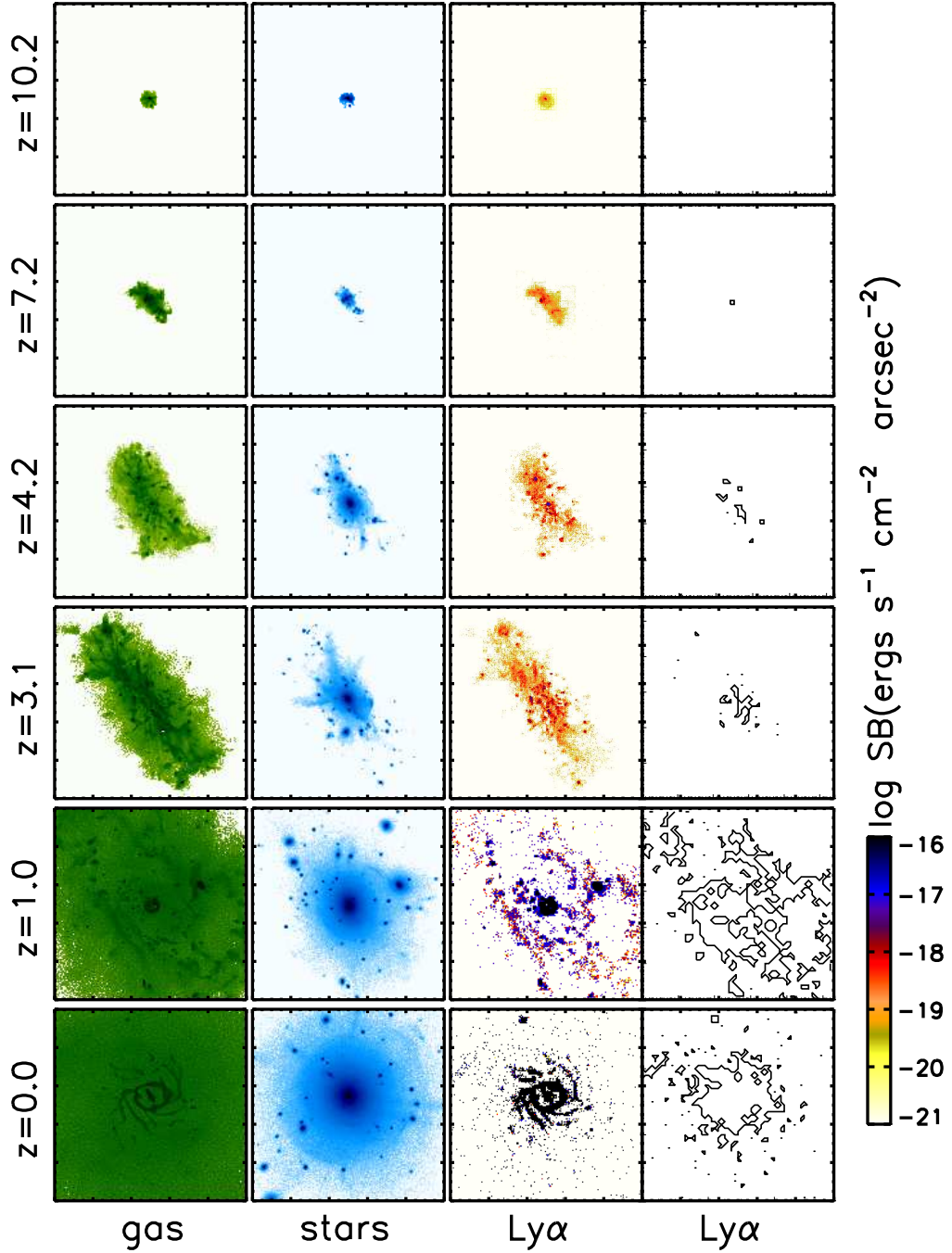}
\caption{
Evolution of the $\lya$ surface brightness of the MW galaxy with redshift. The first and second columns show the distribution of gas (in green) and stars (in blue) from the cosmological simulation, respectively. The third column shows the surface brightness of $\lya$ in log scale with units of $\rm ergs \; s^{-1} \; cm^{-2} \; arcsec^{-2}$. The box size is 200 kpc in physical coordinates. Note the galaxy at $z=0$ is artificially placed at $z=0.1$ in order to obtain the luminosity distance. The right column shows the contour of $\lya$ surface brightness at the level of $10^{-18}~\rm ergs \; s^{-1} \; cm^{-2} \; arcsec^{-2}$, similar to the threshold of recent observations of extended $\lya$ sources by \cite{Matsuda11}. The meshes are coarse grained to be 4 kpc in physical scale to make the smooth contours.
}
\label{fig:img}
\end{center}
\end{figure*}

\begin{figure}
\begin{center}
\includegraphics[scale=0.45]{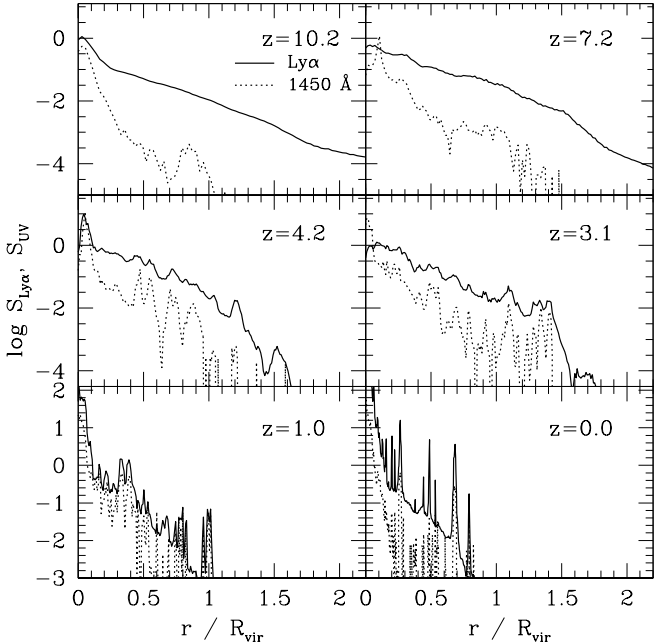}
\caption{
Radial distribution of the surface brightness of the MW galaxy at different redshift. The solid line shows the $\lya$ surface brightness in units of  $10^{-18}\rm~ergs \; s^{-1} \; cm^{-2} \:  arcsec^{-2}$, while the dotted line shows the surface brightness of UV continuum at $1450~\A$ in rest frame in units of $10^{-31}~\rm ergs \; s^{-1} \; cm^{-2} \; Hz^{-1} \; arcsec^{-2}$. 
The surface brightness distribution is derived from the projected flux images in figure~\ref{fig:img} by taking the mean surface brightness as a function of the distance. 
}
\label{fig:uvdist}
\end{center}
\end{figure}

\begin{figure*}
\begin{center}
\includegraphics[scale=1.0]{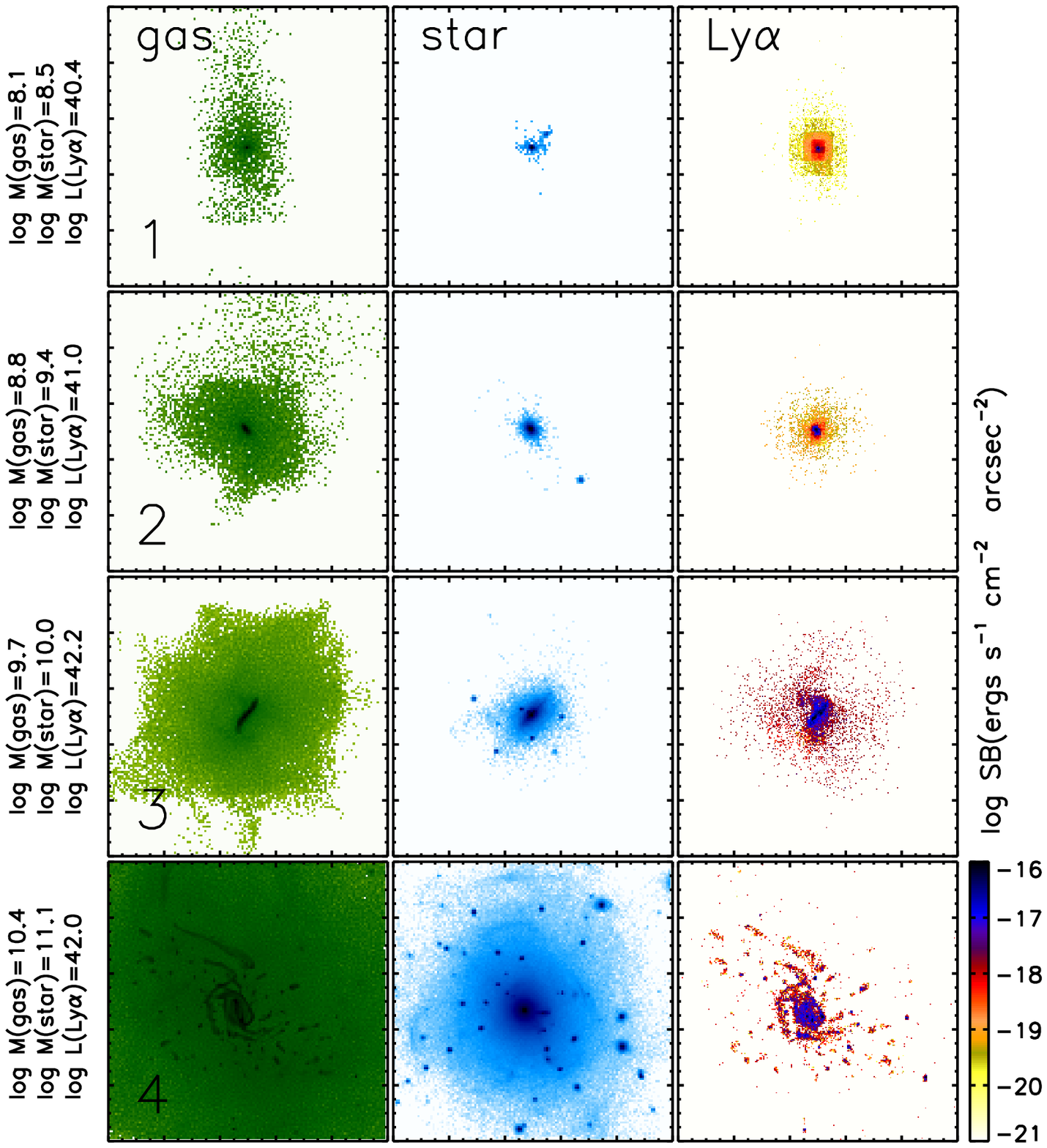}
\caption{The relation between $\lya$ morphology and galaxy mass, exemplified by a sample of four galaxies at $z=0.2$ with different masses, $\Mtot = 4.2\times 10^{9}, 1.7 \times 10^{10}, 1.2 \times 10^{11}$ and $1.6 \times 10^{12}~\Msun$, respectively. The left and middle columns show the distribution of gas (in green) and stars (in blue) from the cosmological simulation, respectively. The right column shows the $\lya$ surface brightness of the corresponding galaxies, with resulting luminosity $\La = 2.6\times 10^{40}, 9.8 \times 10^{40}, 1.8 \times 10^{42}$ and $9.8 \times 10^{41}~\ergs$, respectively. The box size is 250 kpc in physical scale. Note the $\lya$ luminosity is computed by collecting all escaped photons.
}
\label{fig:lyamulti}
\end{center}
\end{figure*}

Figure~\ref{fig:img} shows the $\lya$ surface brightness of the MW galaxy at different redshifts, contrasted with distributions of gas and stars of the galaxy. To facilitate comparison with observations, we adopt an intensity threshold, $S_{\rm Ly\alpha} = 10^{-18}~\rm ergs \; s^{-1} \; cm^{-2} \; arcsec^{-2}$, from a recent survey of extended $\lya$ sources at $z \sim 3$ by \cite{Matsuda11} to show the contours. The $\lya$ distribution appears to trace that of the gas. At $z \ge 7.2$, the galaxy is small and compact, and the $\lya$ emission is confined in the central high-density region. As the galaxy grows in mass and size, the $\lya$ emission becomes more extended. At $z=4.2-3.1$, the gas structure is irregular due to infall along with filament of the main halo. At $z \sim 0.0$, the galaxy shows a disk geometry with spiral structures. Indeed, the $\lya$ map shows filamentary structures at high redshift, and spirals at $z \sim 0.0$. We note that some of the extended $\lya$ sources in the recent observations by \cite{Matsuda11}, which are called $\lya$ blobs, show filamentary structures. Our galaxy at $z =3.1$ has a $\lya$ distribution of $\sim 50$ kpc with a surface brightness above the observational threshold. However, the size is smaller than the observed giant $\lya$ blobs of $\gtrsim 100$ kpc. Such large blobs are probably produced by systems of $M_{\rm halo} \gtrsim 10^{12}~\Msun$, more massive than our model. 
In addition, it was suggested that extended $\lya$ sources of $\gtrsim 40~\rm kpc$ become rare at $z < 1$ \citep{Keel09}.
Recent UV surveys detected $\lya$ from local star-forming galaxies at $z \lesssim 1$, which showed $\sim 10$ kpc $\lya$ distribution above the threshold of $S_{\rm Ly\alpha} \sim 10^{-14}~\rm ergs \; s^{-1} \; cm^{-2} \; arcsec^{-2}$ \citep{Hayes07, Ostlin09}. 
With such a detection sensitivity, our model galaxies at $z \lesssim 1$ show the size of $\lesssim 20$ kpc, in broad agreement with observations.  

To examine the difference in distribution between stars and $\lya$ emission more quantitatively, Figure~\ref{fig:uvdist} shows the surface brightness of UV continuum, which traces young stars, and $\lya$ as a function of the distance from galaxy center at different time. At high redshift ($z \gtrsim 5$), the $\lya$ distributes more extendedly than the stars, as the UV continuum decreases steeply around the virial radius, but at $z < 3$, both emissions appear to have similar radial distribution. Such a transition is mainly due to the difference in $\lya$ production at different epochs. The $\lya$ emission is dominated by collisional excitation, which depends strongly on the gas density, at $z \gtrsim 5$. At a later time, $\lya$ from recombination of ionized gas by stellar radiation becomes more important, so $\lya$ emission follows that of stars.

It was shown by \citet{Cowie11} that nearby LAEs ($z \lesssim 1.0$) have a variety of morphologies, some are disky, while others are mostly compact galaxies. Galaxy morphology is closely tied to galaxy formation and evolution, and it is related to the galaxy mass. As demonstrated in Figure~\ref{fig:lyamulti}, which shows a sample of four galaxies of different masses at $z=0.2$, the $\lya$ morphology changes with galaxy mass. In galaxies with lower mass ($\Mtot \lesssim 10^{11}~\Msun$), the $\lya$ luminosity is low ($\La \lesssim 10^{42}~\ergs$), the $\lya$ morphology is highly compact. At higher mass ($\Mtot \gtrsim 10^{11}~\Msun$), the $\lya$ luminosity is high ($\La \gtrsim 10^{42}~\ergs$), $\lya$ morphology shows disky and spiral structures. This plot suggests that the various morphologies observed in low-redshift LAEs may reflect a wide range of galaxy mass in the sample. 

We note that in Figure~\ref{fig:uvdist}, the azimuthally averaged surface brightness at $z \gtrsim 3$
is somewhat fainter than the detection threshold of recent narrow-band surveys, $\sim 10^{-18} ~\rm ergs \; s^{-1} \; cm^{-2} \; arcsec^{-2}$ \citep{Ouchi08}. However, since the local $\lya$ distribution is inhomogeneous and anisotropic, as shown in Figure~\ref{fig:img}, so bright regions with flux above the threshold may be detectable by such surveys. 

The detectability of these galaxies depends strongly on the sensitivity of the surveys. In the present work, unless noted  otherwise, the $\lya$ luminosity is computed by collecting all escaped photons without a flux limit. If a detection limit of a given instrument is imposed, the luminosity of individual galaxies, in particular that of the faint ones, may be reduced significantly, as suggested by \citet{Zheng10}. 

\subsection{Evolution of Spectral Energy Distribution}

\begin{figure}
\begin{center}
\includegraphics[scale=0.45]{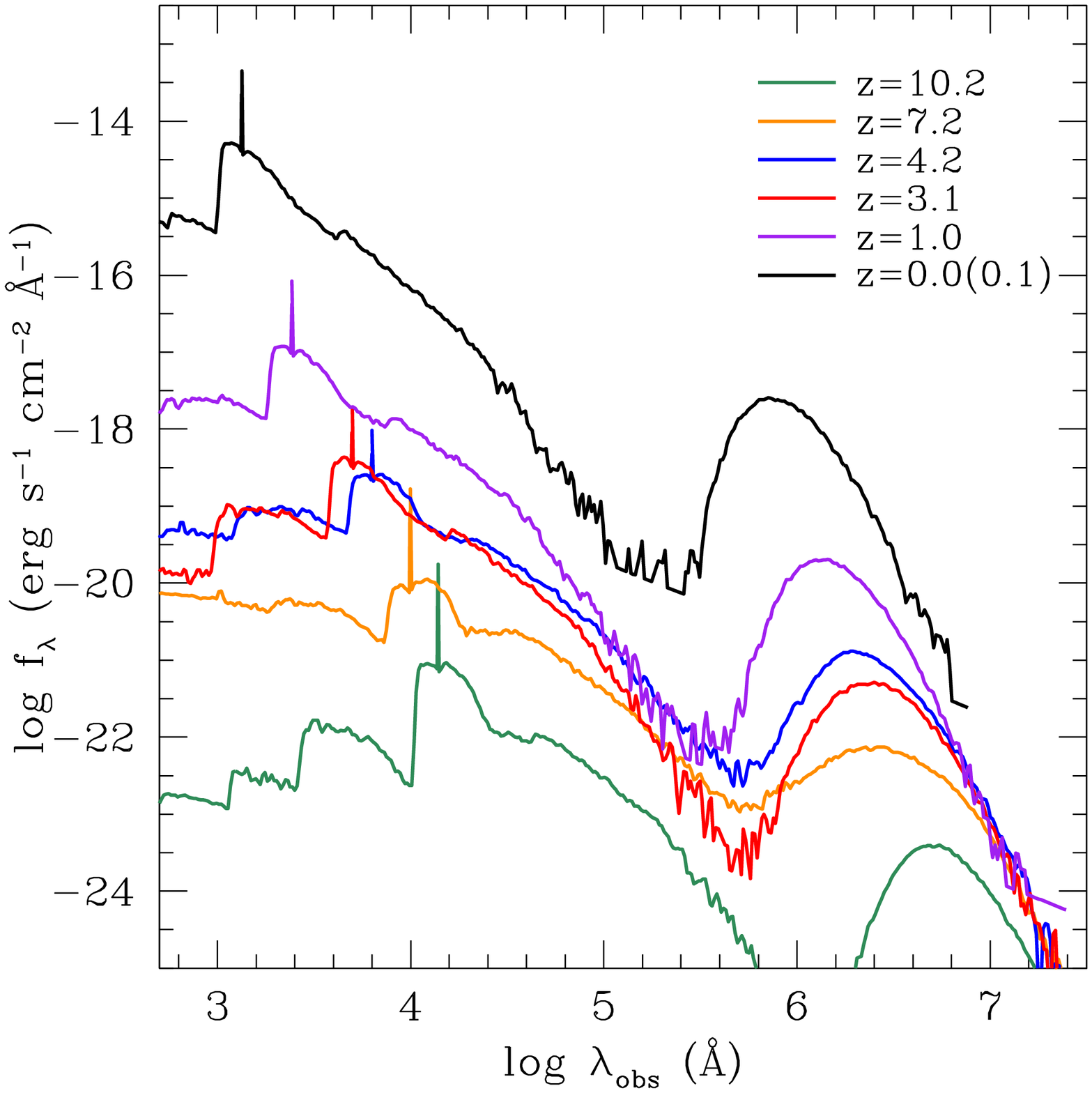}
\caption{
The SEDs of the most massive galaxy in each snap shot.
Only the galaxy at $z=0.0$ is artificially set at $z = 0.1$,
i.e., the flux is estimated by using the luminosity distance of $z = 0.1$.
}
\label{fig:sed}
\end{center}
\end{figure}

The corresponding multi-wavelength SEDs of the galaxy sample in Figure~\ref{fig:img} are shown in Figure~\ref{fig:sed}. The shape of the SED changes significantly from $z = 10.2$ to $z = 0$, as a result of changes in radiation source and environment, since the radiation from stars, absorption of ionizing photons by gas and dust, and re-emission by the dust evolve dynamically with time. The $\lya$ line appears to be strong in all cases. The deep decline of Lyman continuum ($l \le 912~\A$) at high redshifts ($z \gtrsim 7$) is caused by strong absorption of ionizing photons by the dense gas. Galaxies at lower redshift have a higher floor of continuum emission from stars and accreting BHs, a higher ionization fraction of the gas, and a higher infrared bump owing to increasing amount of dust and absorption. Moreover, due to the effect of negative k-correction, the flux at observed frame $\lambda \gtrsim 500~\rm \mu m$ stays close in different redshifts. 

Our calculations show that the main progenitor has a flux of $f_{\nu} = ~0.057~\rm mJy$ at $z \sim 6$ and $0.02~~\rm mJy$ at $z \sim 8.5$ at $850~\rm \mu m$ in observed frame. The new radio telescope, {\it Atacama Large Millimeter/submillimeter Array} (ALMA) may be able to detect such galaxies at $z \sim 6$ with $\sim 2$ hours integration, and at $z \sim 8.5$ for $\sim 20$ hours with 16 antennas \citep{Yajima11A}.

\subsection{The $\lya$ Properties}

\begin{figure}
\begin{center}
\includegraphics[scale=0.5]{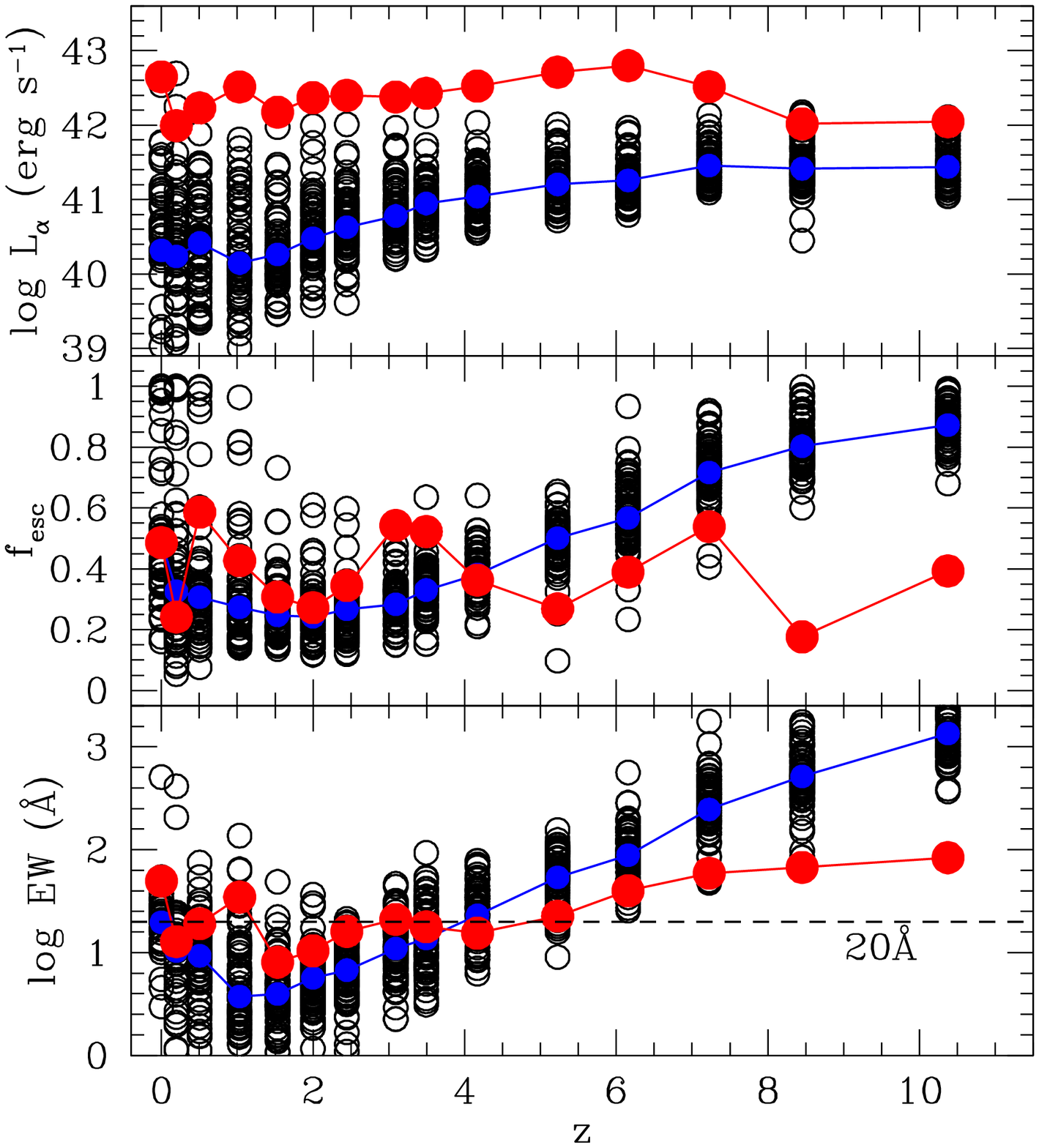}
\caption{
Evolution of the $\lya$ properties of the 60 most massive progenitors from selected redshifts. From top to bottom panel is the emergent unfiltered $\lya$ luminosity (computed by collecting all escaped photons), escape fraction of $\lya$ photons over whole solid angle, and equivalent width in rest frame, respectively. The red filled circle represents the value of the main progenitor at each redshift, while the  blue filled circles indicate the median value of the sample galaxies. 
}
\label{fig:La}
\end{center}
\end{figure}


The resulting $\lya$ properties of the 60 most massive progenitors from selected snapshots, and their evolution with redshift are shown in Figure~\ref{fig:La}. The top panel shows the emergent $\lya$ luminosity $\La$. The main progenitor has a luminosity of $\La \sim 10^{42}~\ergs$ at $z \lesssim 2$, then increases to $\La \sim 10^{43}~\ergs$ at $z = 2-6$, owing to the increase of SFR. At high redshift $z \gtrsim 6$, the $\La$ decreases to $\sim 10^{42}~\ergs$ due to low escape fraction and absorption by dust. 
At high redshift, the galaxy size is small and most of the stars form around the galaxy center. Hence, the dust compactly distributes around young stars, which effectively absorbs the ionizing photons. As a result, the intrinsic $\lya$ emissivity drops even though SFR is enhanced by the accretion of cold gas.  However, most of other model galaxies show that the escape fraction of $\lya$ and UV continuum photons monotonically increases with redshift because of lower dust content. The escaping process of continuum photons will be discussed in detail in a forthcoming paper (Yajima et al. in preparation).

Most of the galaxies at $z \gtrsim 3$ in this simulation have a luminosity below $10^{42}~\ergs$, the lower limit of many current observations using narrow-band filters, so they may not be observable. However, the improved sensitivity of deep survey of \citet{Cassata11} reaches $\La \sim 10^{41}~\ergs$ at $z \sim 2 - 6.6$, which may detect more faint galaxies as the ones in our sample. As mentioned in previous sections, the luminosity is calculated as the sum of all escaped photons. If we consider only pixels brighter than $S = 10^{-18} ~\rm ergs \; s^{-1} \; cm^{-2} \; arcsec^{-2}$, the detection threshold of high-redshift LAE surveys \citep[e.g.,][]{Ouchi08, Ouchi10}, the total flux of the main progenitor is reduced by a factor of a few, resulting in $\La \sim 1- 2 \times 10^{42}~\ergs$ in redshift  $z \sim 2 - 6$, which is close to the observed $L^{*}_{\lya}$ of LAEs in this redshift span \citep[e.g.,][]{Gawiser07, Gronwall07, Ouchi08, Ciardullo11}. This main progenitor has a halo mass of $\sim 10^{11}~\Msun$ at $z \sim 6$, and $\sim 10^{12}~\Msun$ at $z \sim 2$, in good agreement with suggestions from clustering analysis by \citet{Ouchi10}. These results suggest that some the observed LAEs at $z \sim 2 - 6$ may be similar to the main progenitors of MW-like $L^*$ galaxies at high redshifts.


The calculated escape fractions of $\lya$ photons ($\fesc$) are shown in the middle panel of Figure~\ref{fig:La}. 
Here, the $\fesc$ is estimated by correcting all escaped photons over whole solid angle and dividing by intrinsically emitted photon number. Unlike the SFR, the $\fesc$ has higher values at lower redshift $z \lesssim 2$, then decreases gradually to $\sim 20~\%$. At $z \gtrsim 4$ the $\fesc$ increases again. The $\fesc$ of the main progenitor fluctuates in the range of $\sim 20 - 60 ~\%$. The median $\fesc$ at $2 \lesssim z \lesssim 4$ is $\sim 30~\%$, which is consistent with the recent observation by the HETDEX pilot survey \citep{Blanc11}. At lower redshift $z \lesssim 1$, there is a large dispersion in $\fesc$, similar to the recent observation by \cite{Atek09}. This large scattering may be caused by variation in a number of physical properties such as SFR, metallicity, and disk orientation. We will discuss the dependence of $\fesc$ on these properties in detail in Section~\ref{sec:fesc}.

We note that the $\lya$ RT calculations in our work, which take into account local ionization structure and inhomogeneous density distribution of gas and dust, produce a smaller escape fraction ($\fesc \sim 20 - 80~\%$ at $z \sim 6$) than that in previous semi-analytical work of \citet[]{Salvadori10} ($\fesc \gtrsim 80 \%$) and \citet[]{Dayal11b} ($\fesc \sim 60 - 90~\%$), in which a uniform slab model was assumed. We find that more than half of $\lya$ photons can be absorbed, because dense gas and dust around the star-forming and $\lya$-emitting regions absorb the photons effectively.


The EW of $\lya$ line is defined by the ratio between the $\lya$ flux and the UV flux density $f_{\rm UV}$ in rest frame, where the mean flux density of $\lambda = 1300 - {1600} \; \A$ in rest frame is used. The resulting $\lya$ EWs are shown in the bottom panel of Figure~\ref{fig:La}. Most of the galaxies have $\EW \gtrsim 20\; \A$, they are therefore classified as LAEs \citep[e.g.,][]{Gronwall07}. The median EW increases with redshift, from $\sim 30\; \A$ at redshift $z =0$ to $\sim 820\, \A$ at  $z \sim 8.5$. This trend is in broad agreement with observations that galaxies at higher redshifts appear to have higher EW than their counterparts at lower redshifts \citep[e.g.,][]{Gronwall07, Ouchi08}. The high EW at $z \gtrsim 6$ is produced by excitation cooling, which enhances the $\lya$ emission at high redshift, but at low redshift it reduces the EW as the stellar population ages \citep[e.g.,][]{Finkelstein09}. Recent observations of LAEs at $z \sim 0.2 - 0.4$ shows that most local LAEs, unlike those at $z \gtrsim 3$, have EWs less than $100~\A$ \citep{Deharveng08, Cowie11}, consistent with the trend seen in our model.

We should point out that the results presented in Figure~\ref{fig:La} are ``unfiltered'' by detection limit, and that we caution against taking these numbers too literally when compared with a particular survey, because the observed properties depend strongly  on the observational threshold. Note also in the current work, we did not include the transmission in intergalactic medium (IGM). The $\lya$ properties can be changed by IGM extinction. The neutral hydrogen in IGM at high redshift can scatter a part of $\lya$ photons, and decrease the $\La$ and EW. For example, \citet{Laursen11} suggested that the IGM transmission could be $\sim 20~\%$ at $z=6.5$. The transmission depends sensitively on the viewing angle and the environments of a galaxy, as it is affected by the inhomogeneous filamentary structure of IGM.

\subsection{The $\lya$ Line Profiles}

\begin{figure}
\begin{center}
\includegraphics[scale=0.4]{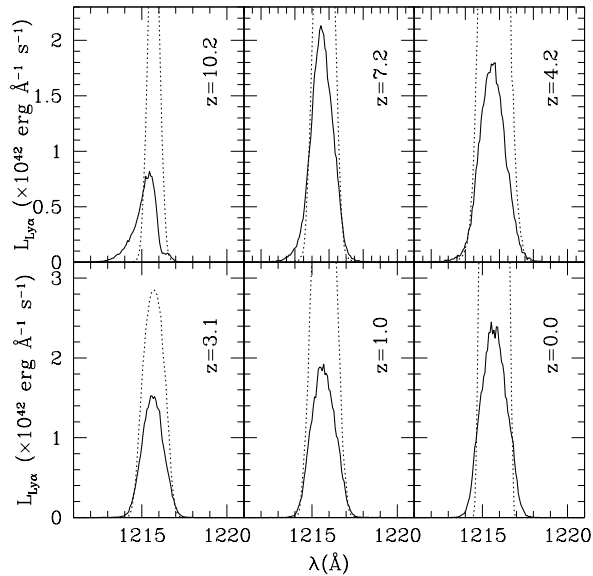}
\caption{
The $\lya$ line profile of the main progenitor at different redshifts.
The dotted- and solid lines are the intrinsic and emergent $\lya$ profile, respectively.
}
\label{fig:profile}
\end{center}
\end{figure}

The emergent $\lya$ emission line of the main progenitor is shown in Figure~\ref{fig:profile}. The frequency of the intrinsic $\lya$ photon is sampled from a Maxwellian distribution with the gas temperature at the emission location in the rest flame of the gas. 
Our sample $\lya$ lines show mostly single peak,
common profiles of LAEs observed both at high redshift ($z \sim 6$) \citep[e.g.,][]{Ouchi10} and in the nearby universe \citep[e.g.,][]{Cowie10}. 

In a static and optically thick medium, the $\lya$ profile can be double peaked, but when the effective optical depth is small due to high relative gas speed or ionization state, there might be only a single peak \citep{Zheng02}. In our case, the flow speed of gas is up to $\sim 300$ km/s, and the gas is highly ionized by stellar and AGN radiation, which result in a single peak. 

In the case at high redshift $z \gtrsim 6$, the gas is highly concentrated around the galaxy center, hence they become optically thick and cause the $\lya$ photons to move to the wing sides. In addition, the profile at $z=10.2$ shifts to shorter wavelength, and it shows the characteristic shape of gas inflow \citep{Zheng02}. Although our simulation includes feedback of stellar wind similar to that of \citet{Springel05d}, the $\lya$ line profile indicates gas inflow in the galaxy. Our result suggests that high-redshift star-forming galaxies may be fueled by efficient inflow of cold gas from the filaments. We will study this phenomenon in detail in Yajima et al. (in preparation).
On the other hand, it was suggested that the asymmetrically shifted profile to the red wing in some LAEs can be made by outflowing gas distribution \citep[e.g.,][]{Mas-Hesse03}. The growing hot bubble gas around star-forming region from supernovae or radiative feedback can cause outflowing, neutral gas-shells, which result in red-shifted line profiles.

Recently, \citet{Yamada12} observed a sample of 91 LAEs at $z = 3.1$, about half of which show double peaks of strong-blue and weak-red features thought to be caused by gas outflow, while others show a symmetric single peak in which the flux ratio of blue wing to red one is about unity. While our model may explain the latter, the missing outflow features in our line sample is probably due to the limitations of our current simulations, e.g., insufficient spatial resolution and simplified treatment of supernovae feedback.

In addition, the line profiles of galaxies at high redshift $z \gtrsim 6$ may be highly suppressed and changed by scattering in IGM \citep[e.g.,][]{Santos04, Dijkstra07, Zheng10, Laursen11}, because the $\lya$ transmission through IGM is very low at the line center and at shorter wavelengths by the Hubble flow \citep[e.g.,][]{Laursen11}.
Even at lower redshift $z \sim 3$, the optical depth of IGM can be high depending on the viewing angle and the location  of the galaxy \citep[e.g.,][]{Laursen11}. 
Therefore, the $\lya$ flux with inflow featured in our model galaxies may be suppressed and the shape may change to a single peak with only the red wing, or a double peak with strong-red and weak-blue as in Figure~7 of \citet{Laursen11}.

%
%
\section{Discussion}

\subsection{Dependence of $\lya$ Properties on Galaxy Properties}
\label{sec:fesc}

\begin{figure*}
\begin{center}
\includegraphics[scale=0.9]{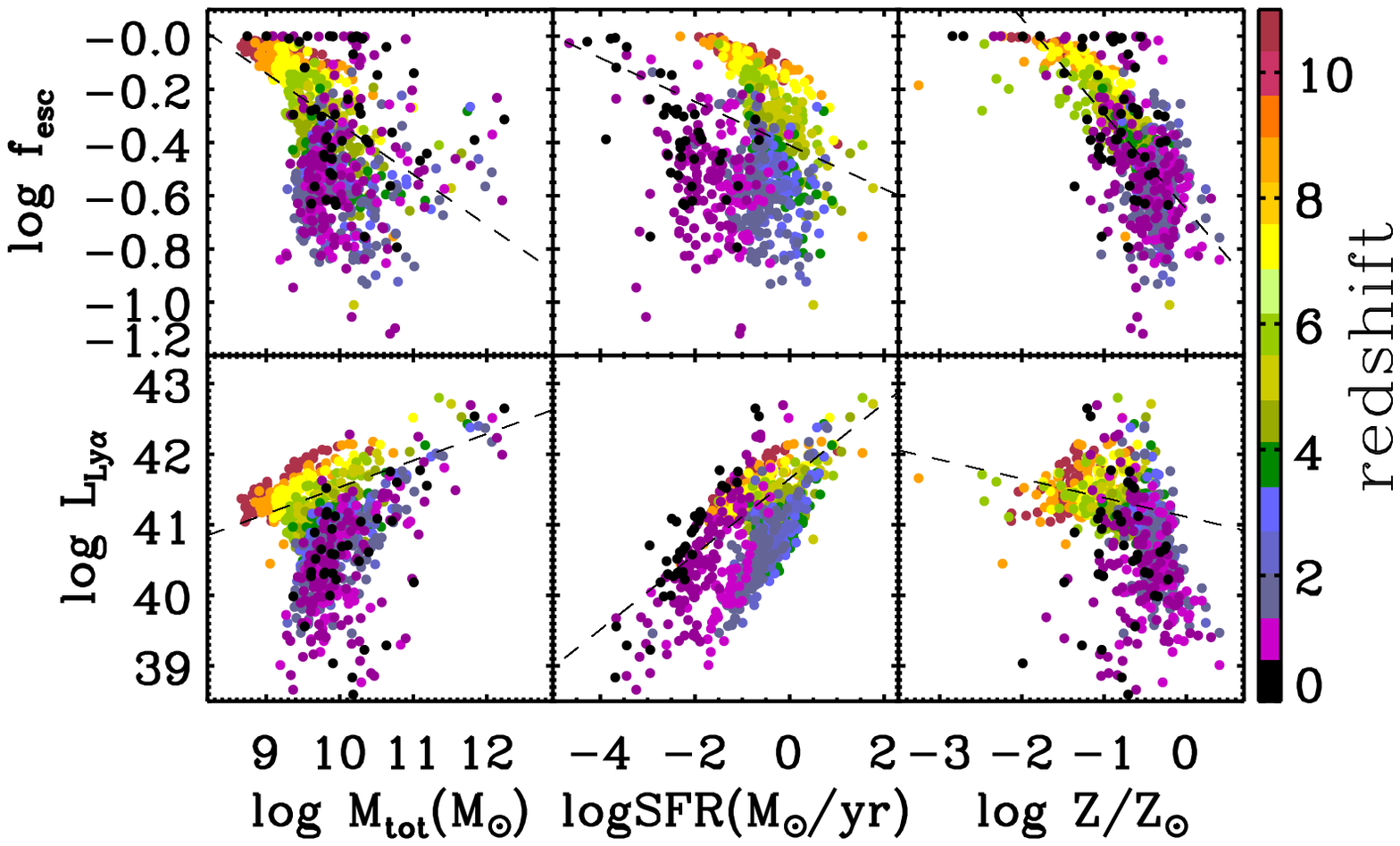}
\caption{Dependence of $\lya$ properties on various physical properties of a galaxy. 
{Upper~:} The relation between $\lya$ escape fraction and halo mass, 
SFR and metallicity. The different color represents the different redshift, as indicated in the color bar.
The dash lines are least-absolute-deviations fittings. {Lower~:} The relation between $\lya$ luminosity and halo mass, 
SFR and metallicity.
}
\label{fig:fesc_all}
\end{center}
\end{figure*}

\begin{figure}
\begin{center}
\includegraphics[scale=0.4]{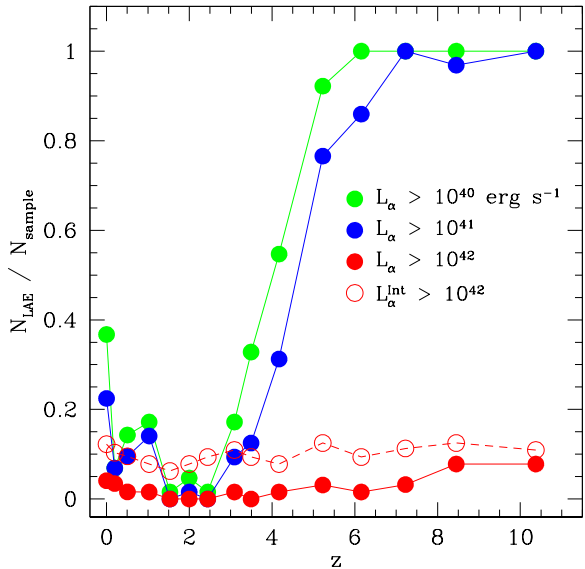}
\caption{
Number fraction of LAEs in our galaxy sample.
The green, blue and red filled circles represent
the fraction at $\lya$ luminosity threshold,
$10^{40}, ~10^{41}$, and $10^{42}~\ergs$, respectively. 
}
\label{fig:frac}
\end{center}
\end{figure}

As shown in previous Sections, the $\lya$ properties vary significantly in different galaxies. Here we explore the dependence on a number of physical properties of a galaxy. Figure~\ref{fig:fesc_all} shows the dependence of escape fraction $\fesc$ (top panels) and $\lya$ luminosity $\La$ (bottom panels) on the galaxy mass, SFR, and metallicity Z. We apply a least-absolute-deviations fitting to the data using a power-law function, ${\rm log} Y = \alpha {\rm log}X + \beta$.

The mass dependence of $\fesc$ has a large dispersion, but from our fitting, $\alpha \sim - 0.02$ and $\beta \sim -0.17$, which suggests that $\fesc$ roughly decreases with the total mass, consistent with the results of \citet{Laursen09b}. At $M \sim 10^{10-11}~\Msun$, the $\fesc$ is mostly constant at $\sim 10 - 30~\%$. In contrast, $\fesc$ is more tightly correlated with the SFR, with $\alpha \sim -0.08$ and $\beta \sim -0.41$. At high SFR, dust  can be enriched quickly by type II supernovae, and can effectively absorb the $\lya$ photons. In addition, galaxies with high SFR have more hydrogen gas. The gas decreases the mean free path of $\lya$ photons, resulting in the increase of the dust optical depth which reduces the escape fraction. In addition, the $\fesc$ decreases with metallicity, $\alpha \sim -0.35$ and $\beta \sim -0.65$. Since the dust content linearly increases with metallicity in our model, the $\lya$ photons can be absorbed effectively by gas with high metallicity. This trend is consistent with observational indication by \cite{Atek09} and \cite{Hayes10}. 

On the other hand, the luminosity $\La$ has different relationships with these properties from the $\fesc$. The $\La$ is also roughly correlated with the mass, $\La \simeq 10^{37.7}\times\Mtot^{0.38}$, with a large dispersion. Only massive galaxies with $\Mtot \gtrsim 10^{11}~\Msun$ have the $\lya$ luminosity of $\La \ge 10^{42}~\ergs$. This is consistent with suggestions from clustering analysis of observed LAEs at $z=2-3$ \citep[e.g.,][]{Gawiser07, Guaita10}. In our model, the massive galaxies at $z=2-3$ evolve into $L^{*}$ galaxies at $z=0$. Hence, our results support the suggestion by \cite{Gawiser07} that the observed LAEs with $\La \gtrsim 10^{42}~\ergs$ at $z=2-3$ are likely progenitors of local $L^{*}$ galaxies. 

The $\La$ has the tightest correlation with SFR among the properties investigated here: $\La \simeq 10^{41.7}\times \rm SFR^{0.53}$ In the literature, a simple linear relation is commonly used, with $\La \;(\ergs) = 1.1 \times 10^{42}\times \rm{SFR} \;(\Msunyr)$, assuming that $\La / L_{\rm H\alpha} = 8.7$ (case B). However, our result suggests that the relation between $\La$ and SFR becomes somewhat shallower due to the dependence of $\fesc$ on SFR. Finally, the emergent $\La$ does not show a strong dependence on metallicity, $\La \simeq 10^{41.1}\times (Z/Z_{\odot})^{-0.27}$. This is due to the fact that, although the intrinsic $\La$ increases with halo mass (so does SFR and metallicity), the $\fesc$ decreases with metallicity, so $\La$  of higher-metallicity galaxies is suppressed by dust absorption. 

We should point out that the large scatter in the correlations in Figure~\ref{fig:fesc_all} may be due to the small volume of our simulation and the small number of our galaxy sample. In addition, as we discuss in Section~\ref{sec:limit}, a number of limitations of our model, such as the simplified ISM model and insufficient resolutions, may contribute to uncertainty in these relations. Moreover, the luminosity scaling relations may change under some specific detection limits. We will study these relations in detail with improved model and simulations in future work.

\subsection{Redshift Dependence of LAE Fraction}

The number fraction of LAEs ($\fa$) in our sample is shown in figure~\ref{fig:frac}.
The detection limit of $\lya$ varies in different surveys. At high redshifts ($z \gtrsim 3$), the LAE detection in most of observations has been confined to $\La \gtrsim 10^{42}~\ergs$. Here, we derive the $\fa$ with three $\La$ thresholds, $\La \gtrsim 10^{40}, ~10^{41}, ~10^{42}~\ergs$ with EW of $\gtrsim 20~\A$. The $\fa$ with $\La \gtrsim 10^{40}, 10^{41}~\ergs$ rapidly increases from $z=0$ to $\sim 5$,
and then remains nearly constant with higher values $\gtrsim 0.8$ at $z \gtrsim 5$.
The trend is roughly similar to the SFR history (Figure~\ref{fig:sfr}).
Since the $\La$ is tightly correlated with the SFR (Figure~\ref{fig:fesc_all}),
the number of galaxies with $\La \gtrsim 10^{40}, ~10^{41}$ increases at $z \sim 0 - 4$.
On the other hand, although the SFR decreases at $z \gtrsim 4$,
the $\fesc$ increases due to low metallicity.
Hence, the $\fa$ does not decrease at $z \gtrsim 4$.

Meanwhile, the $\fa$ with $\La \gtrsim 10^{42}~\ergs$ is nearly constant, and shows $\sim 2 - 10~\%$.
Since the SFR tightly correlates with $\La$, and it roughly increases with the galaxy mass, some massive galaxies can be LAEs with $\La \gtrsim 10^{42}~\ergs$. In addition, the $\fa$ of LAEs having intrinsic $\La \gtrsim 10^{42}~\ergs$ change with cosmic star formation history. However, the $\fesc$ decreases around the phase of SFR peak, and therefore suppresses the $\fa$.

On the other hand, at lower redshift, the observations indicate that number density of LAEs decreases by some factors 
\citep[e.g.,][]{Cowie10}. The discrepancy may come from the difference in density field and the small box in our simulation. Our initial condition is a somewhat special one which is focused on a MW-size galaxy, and the zoom-in simulation region is $\sim 5^3~ h^{-3} \Mpc^{3}$. Therefore, our simulation cannot reproduce the global statistics in observations.  
Moreover, the LAEs fraction having $EW > 25~\A$ in this work shows $\sim 44\; \%$ at z=4 and $\sim 100\; \%$ at $z = 6$, which is somewhat higher than the LAE fraction in LBG sample \citep{Stark10, Stark11, Pentericci11, Schenker12, Ono12}.
However, in observation, the LAE fraction increases with decreasing UV brightness.
Most of our model galaxies at $z \gtrsim 3$ are fainter than the detection threshold in the LBG observation.
Since the number of galaxies brighter than the threshold of LBG observation is quite small (less than ten), we need a larger sample covering a wide mass range to verify the model of LAEs. In addition, although some LAEs have been observed with UV continuum, and hence categorized as LBGs, it is inadequate to study LAEs from LBG-only sample, because a large fraction of LAEs may have UV continuum under the detection limit of current observations. We will address the general properties such as luminosity function, EW distribution and clustering systematically by using a set of uniform simulations with mean density field in larger volumes in future work.

\subsection{The Viewing-angle Scatter of Escaping $\lya$ Photons}
\label{sec:angle}

\begin{figure}
\begin{center}
\includegraphics[scale=0.45]{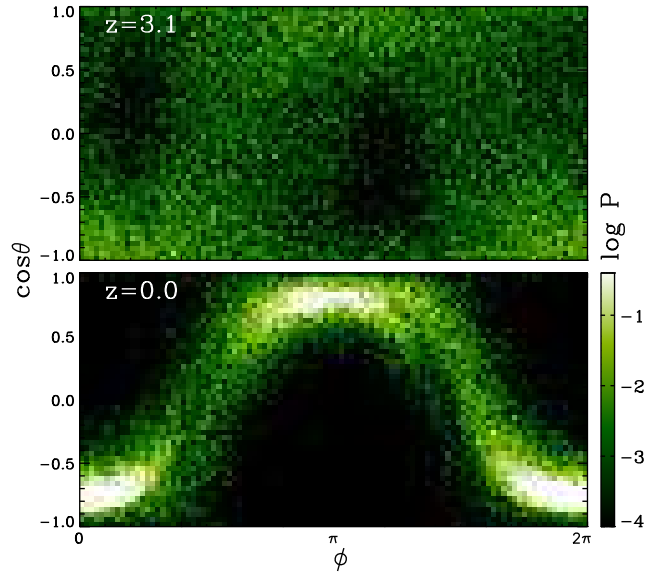}
\caption{The ``viewing-angle scatter'' 
of escaping $\lya$ photons depends strongly on galaxy morphology and orientation. Shown here is the $\lya$ escape probability in a irregular galaxy progenitor at z=3.1 (top panel) and the spiral MW galaxy at z=0 (bottom panel). The color bar indicates the probability per unit  $\rm d\phi, dcos\theta$. There is no clear direction in the irregular galaxy, but in the disk galaxy, the $\lya$ photons escape in a preferred direction normal to the disk.
}
\label{fig:fescmap}
\end{center}
\end{figure}

Despite their high metallicity, a fraction of galaxies at low redshift $z \lesssim 1$ show high escape fraction $\fesc$ of $\lya$ photons (Figure~\ref{fig:fesc_all}). We find that the escaping angle of the $\lya$ photons depends strongly on the galaxy morphology and orientation, a phenomenon we dub as the ``viewing-angle scatter''. 
Disky objects seen edge on can be hundred times fainter than  the same objects seen face on.
In a galaxy which has a gas disk, the $\lya$ photons escape in a preferred direction normal to the disk, but there is no clear escaping direction in compact or irregular galaxies without a gas disk. We demonstrate this effect in Figure~\ref{fig:fescmap}. We first estimate the normal direction to the gas disk according to the total angular momentum of the gas, and set $\theta = 0^{\circ}$ along this direction. In a galaxy with irregular morphology such as the main progenitor at $z = 3.1$, there is no clear preferred escaping angle, as illustrated in the top panel of Figure~\ref{fig:fescmap}. However, in a spiral galaxy with rotationally supported gas disk such as the MW galaxy at z=0 in our simulation, the escaping angle is strongly confined to $\rm{cos\theta \simeq \pm 1}$, corresponding to $\theta \simeq 0^{\circ}$ or $180^{\circ}$, as shown in the bottom panel of Figure~\ref{fig:fescmap}. This is due to the fact that the $\lya$ photons have the minimum optical depth along the normal direction to the gas disk. More than $60~\%$ $\lya$ photons escapes to the direction of $\rm |cos~\theta| \lesssim 0.5$. Generally the $\lya$ flux from our model galaxies can scatter around the mean value typically by a factor of ten just from different orientations.

As illustrated in Figure~\ref{fig:img}, most galaxies in our simulation have highly irregular shapes at high redshift due to accretion and gravitational interaction. At z=0, a number of them evolve into spiral disks. The ``viewing-angle scatter'' 
explains why we see high $\lya$ escape fractions in a number of low-z galaxies, and the fact that $\lya$ is detected in a large number of face-on spiral galaxies in the nearby universe \citep[e.g.,][]{Cowie10}.

\subsection{Contribution of Excitation $\lya$ Cooling}
\label{sec:exc}

\begin{figure}
\begin{center}
\includegraphics[scale=0.4]{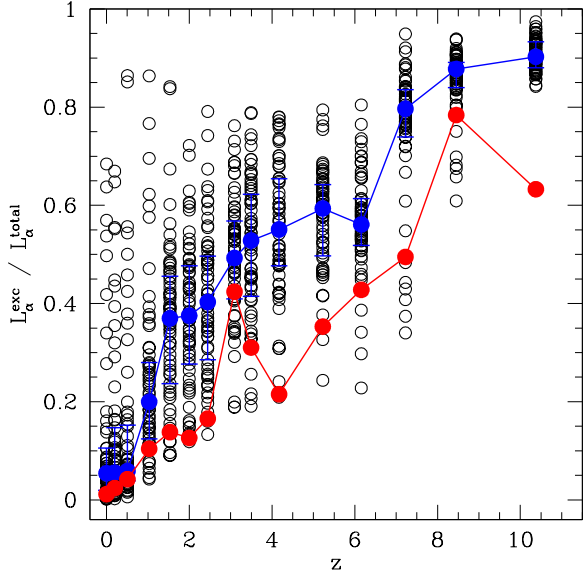}
\caption{
The fraction of excitation cooling $\lya$ to the total intrinsic $\lya$ luminosity as a function of redshift.
The red filled circles indicate the values of the main progenitor, while the blue filled circles represent the median value of the galaxy sample indicated with black open circles.
The error bars show the quartiles.
}
\label{fig:exc}
\end{center}
\end{figure}

There are two major mechanisms to generate $\lya$ emission, the recombination of ionizing photons and the collisional excitation of hydrogen gas. However, the relative contribution between the two mechanisms is not well understood. From our calculations, we find that the contributing fraction of excitation $\lya$ emission to the total intrinsic $\lya$ luminosity increases with redshift, as shown in figure~\ref{fig:exc}. 

In our cosmological simulation, galaxy evolution is accompanied by cold, filamentary gas streams with temperature $T\sim 10^{4-5}~\rm K$, which penetrate deep inside dark matter halos (Zhu et al. in preparation, Yajima et al. in preparation), a phenomenon also reported by other groups \citep{Katz03, Keres05, Keres09, Birnboim03, Dekel06, Ocvirk08, Brooks09, Dekel09}. Such cold gas can efficiently produce the excitation $\lya$ cooling photons \citep{Dijkstra09, Faucher09, Goerdt10}. At higher redshift, galaxies form through more efficient gas accretion and more frequent merging event. As a result, the contributing fraction increases with redshift, and becomes dominant at $z \gtrsim 6$. This excitation mechanism does not depend on the stellar radiation, and can therefore produce high $\lya$ EWs. We find that the EWs of LAEs increases significantly at $z \gtrsim 6$, reaching $\gtrsim 10^3~\A$ at $z \sim 10$. This is larger than the upper-limit of EW, $240~\A$, which considers only stellar sources assuming a Salpeter IMF with solar abundance of metallicity \citep{Charlot93}. Although the upper-limit increases with decreasing metallicity, it was suggested that top-heavy IMF like Pop III stars are needed for making $\rm EW > 400~\A$ \citep{Schaerer03, Raiter10}. However, even though Salpeter-IMF is used in this work and the stellar metallicity is mostly $Z / Z_{\rm \odot} \gtrsim 10^{-3}$, we find that the EW can be higher than the upper-limit by the efficient excitation $\lya$ emission. On the other hand, the $\lya$ line is strongly damped by IGM correction at $z \gtrsim 6$ \citep{Haiman02, Laursen11, Dayal11b}, which can result in a lower EW. The suppression by IGM highly depends on the inhomogeneous ionization structure around LAEs \citep[e.g.,][]{McQuinn07, Mesinger08, Iliev08}. We will address the detectability of high-redshift LAEs and EW after IGM correction by running large-scale $\lya$ RT in IGM in future work.

\subsection{$\lya$ Luminosity Functions}
\label{sec:LF}

The simulation box in this work is too small to study global statistics directly. As a rough estimate, we use  the luminosity -- halo mass correlation we find above may be used to construct $\lya$ luminosity functions (LFs) at different redshift when combined with halo mass functions from large-box, general cosmological simulations. For example, at $z = 3.1$, we divide all galaxies in the snap shot by the halo mass with 0.25 dex, and fit to the median value of each bin, this gives a correlation of $\La ~({\ergs})= 10^{32.94} \times (M_{\rm halo}^{0.79} / \Msun)$. We then use this to convert the halo mass function of \citet{Sheth99} to the $\lya$ LF. 

\begin{figure}
\begin{center}
\includegraphics[scale=0.4]{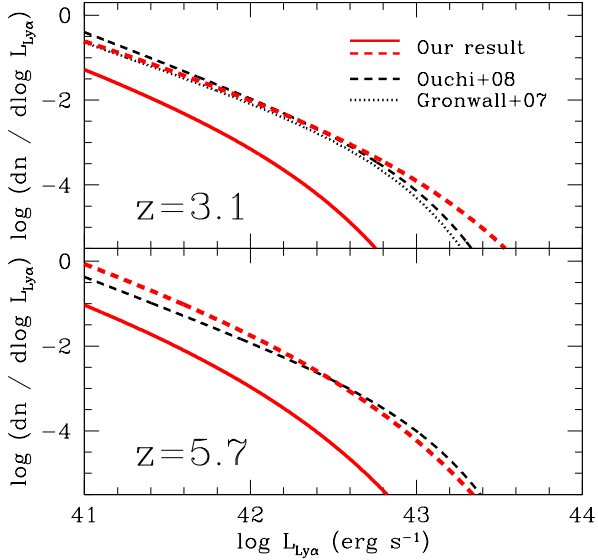}
\caption{
Derived $\lya$ luminosity functions from our model in comparison with observations. The red, solid lines are filtered LFs above a  detection threshold of $S_{\rm Ly\alpha} = 10^{-18}~\rm ergs \; s^{-1} \; cm^{-2} \; arcsec^{-2}$ \citep{Ouchi08}, while the red, dashed lines represent unfiltered ones. The black dashed lines are the Schechter function at $z=3.1$ and $5.7$ derived from LAEs observation by \citet{Ouchi08}, and the black dotted line is from observation at $z=3.1$ by \citet{Gronwall07}.
}
\label{fig:LF}
\end{center}
\end{figure}

Figure~\ref{fig:LF} shows the resulting $\lya$ LFs in comparison with observations at redshift $z = 3.1, 5.7$, respectively. The red solid curves are LFs above a  detection threshold of $S_{\rm Ly\alpha} = 10^{-18}~\rm ergs \; s^{-1} \; cm^{-2} \; arcsec^{-2}$ \citep{Ouchi08, Ouchi10}, while the red dashed lines represent LFs from total luminosity (counting all escaped photons without a flux cut). 

While the un-filtered LFs seem to agree with observations of \cite{Gronwall07} and \cite{Ouchi08}, the filtered ones are significantly off. The difference comes from the reduction of $\La$ and $\fesc$ due to the flux cut. Moreover, the dispersion in the luminosity -- halo mass relation at different redshift may cause a large scatter in the LFs.  This plot suggests that the current simulation in this work is not suitable to study a large galaxy population and its statistical properties, because there are too few observable LAEs. Moreover, as discussed earlier, the predicted $\lya$ properties may be affected by a number of numerical and physical limitations of our model. For example, the one-phase model currently used in the present work may  underestimates the density of cold hydrogen gas, and hence underestimates the $\lya$ flux. We will study the $\lya$ LFs at different redshift in a forthcoming paper with the improved  $\art$ which incorporates a two-phase ISM model, 
and a general simulation with mean overdensity in a larger volume (Yajima et al, in preparation).

\subsection{Limitations of Our Model}
\label{sec:limit}

As demonstrated above, our model is able to explain a number of observed properties of LAEs at different redshift. However, we should point out that our current simulations suffer from a number of major limitations which may affect the predicted $\lya$ properties. 

\begin{itemize} 

\item In the current work, we use a one-phase ISM model, which considers the average density and temperature of the gas. Such a model likely underestimates the density of cold hydrogen gas, which may lead to significant underestimate of the $\lya$ emission coming from cold ($\sim 10^4$~K), dense gas. On the other hand, such a model also underestimates the amount of dust associated with cold molecular gas, which likely results in underestimate of absorption of $\lya$ photons by gas and dust. We will investigate the $\lya$  RT and ionization structures in a two-phase ISM model in a forthcoming paper.

\item The absorption and transmission of IGM are not taken into account in the RT calculations. As discussed in the previous section, these two effects may suppress the $\lya$ flux and change the line profiles. 

\item The simulations do not have sufficient resolutions to resolve dense regions and outflow, which requires a high spatial resolution of  $\sim \rm pc$ \citep[e.g.,][]{Fujita09}. It is a challenge for cosmological simulations to resolve both the inflow gas from large scales of $\sim \rm Mpc$ and the outflow from pc-scale star forming regions. For a simulation with a box of 100 Mpc like the one we have, this requires a large dynamical range over eight orders of magnitude, which is beyond the scope of our current work.

\item The simulation box in this work is too small to study a large galaxy population, as well as effects of environment on galaxy properties and their evolution. One needs uniform simulations in large volumes in order to systematically investigate the formation and evolution of L* galaxies.

\end{itemize}

Finally, we stress once again that caution should be taken when comparing directly the results from our calculations  to data from a given survey, because, as discussed above, the observed $\lya$ properties depend sensitively on a number of factors, including galaxy properties, viewing angle, model parameters, and observational threshold.


%
%
\section{Summary}

To summarize, we have investigated the $\lya$ properties of progenitors of a local $L^{*}$ galaxy 
by combining cosmological hydrodynamic simulations with three-dimensional radiative transfer calculations using the new $\art$ code. Our cosmological simulation follows the formation and evolution of a Milky Way-size galaxy and its substructures from redshift $z = 127$ to $z = 0$. It includes important physics of dark matter, gas dynamics, star formation, black hole growth, and feedback processes, and has high spatial and mass resolutions to resolve a MW-like galaxy at z=0 and its progenitors at higher redshifts. Our radiative transfer couples $\lya$ line, ionization of neutral hydrogen, and multi-wavelength continuum radiative transfer, which enables a self-consistent and accurate calculation of the $\lya$ properties in galaxies.

We find that the main progenitor of the MW galaxy is $\lya$ bright at high redshift, with the emergent $\lya$ luminosity close to the observed characteristic $L_{\lya}^{*}$ of LAEs at $z \sim 2 - 6$. Most of the fainter galaxies in the simulation fall below the detection threshold of many current surveys. The $\lya$ escape fraction correlates with a number of physical properties of the galaxy, such as mass, SFR and metallicity. We find a ``viewing-angle scatter'' in which the photon escape depends strongly on the galaxy morphology and orientation, such that the $\lya$ photons escape in a preferred direction normal to the gas disk in disk galaxies, but randomly in compact or irregular galaxies. Moreover, the EWs of LAEs increases with redshift, from tens of Angstroms at redshift $z \sim 0$ to hundreds of Angstroms at  $z \sim 8.5$. Furthermore, we find that high-redshift LAEs show $\lya$ line profiles characteristic of gas inflow, and that the $\lya$ emission by excitation cooling increases with redshift, accounting $\sim 50 ~\%$ of the total at $z \gtrsim 6$. 

Our results suggest that galaxies at high redshift form through accretion of cold gas, which accounts for the the high EWs, the blue-shifted line profiles, and the dominant contribution from excitation cooling in $\lya$ emission. Moreover, some of the observed LAEs at $z \sim 2-6$ with $\La \sim 10^{42-43}~\ergs$ may evolve into present-day $L^{*}$ galaxies such as the Milky Way.

\acknowledgments

We thank Carlos Frenk for kindly providing us the Aquila initial condition for the cosmological simulation. We thank Mark Dijkstra, Claude-Andr{\'e} Faucher-Gigu{\`e}re, Eric Gawiser, Lars Hernquist, and Avi Loeb 
for stimulating discussions and helpful comments, as well as the referee for an insightful report which has helped improve the mannuscript. Support from NSF grants AST-0965694 \& AST-1009867 (to YL), AST-0807075 (to TA), and AST-0807885 (to CG \& RC) is gratefully acknowledged. YL thanks the Institute for Theory and Computation (ITC) at Harvard University where the project was started for warm hospitality. We acknowledge the Research Computing and Cyberinfrastructure unit of Information Technology Services at The Pennsylvania State University for providing computational resources and services that have contributed to the research results reported in this paper (URL: http://rcc.its.psu.edu). The Institute for Gravitation and the Cosmos is supported by the Eberly College of Science and the Office of the Senior Vice President for Research at the Pennsylvania State University.


\begin{thebibliography}{116}
\expandafter\ifx\csname natexlab\endcsname\relax\def\natexlab#1{#1}\fi

\bibitem[{{Acquaviva} {et~al.}(2011)}]{Acquaviva11}
{Acquaviva}, V., {Vargas}, C., {Gawiser}, E., \& {Guaita}, L.
2011, arXiv:1111.6688

\bibitem[{{Atek} {et~al.}(2009){Atek}, {Kunth}, {Schaerer}, {Hayes},
  {Deharveng}, {{\"O}stlin}, \& {Mas-Hesse}}]{Atek09}
{Atek}, H., et al. 2009,  \aap, 506, L1

\bibitem[{{Barnes} \& {Hut}(1986)}]{Barnes86}
{Barnes}, J., \& {Hut}, P. 1986, \nat, 324, 446

\bibitem[{{Birnboim} \& {Dekel}(2003)}]{Birnboim03}
{Birnboim}, Y., \& {Dekel}, A. 2003, 
  \mnras, 345, 349

\bibitem[{{Blanc} {et~al.}(2011){Blanc}, {Adams}, {Gebhardt}, {Hill}, {Drory},
  {Hao}, {Bender}, {Ciardullo}, {Finkelstein}, {Fry}, {Gawiser}, {Gronwall},
  {Hopp}, {Jeong}, {Kelzenberg}, {Komatsu}, {MacQueen}, {Murphy}, {Roth},
  {Schneider}, \& {Tufts}}]{Blanc11}
{Blanc}, et al. 2011, \apj, 736, 31

\bibitem[{{Bond} {et~al.}(2011){Bond}, {Gawiser}, {Guaita}, {Padilla},
  {Gronwall}, {Ciardullo}, \& {Lai}}]{Bond11}
{Bond}, N., et al. 2011,  ArXiv e-prints

\bibitem[{{Bondi}(1952)}]{Bondi52}
{Bondi}, H. 1952, \mnras, 112, 195

\bibitem[{{Bondi} \& {Hoyle}(1944)}]{Bondi44}
{Bondi}, H., \& {Hoyle}, F. 1944, 
  \mnras, 104, 273

\bibitem[{{Bongiovanni} {et~al.}(2010){Bongiovanni}, {Oteo}, {Cepa}, {P{\'e}rez
  Garc{\'{\i}}a}, {S{\'a}nchez-Portal}, {Ederoclite}, {Aguerri}, {Alfaro},
  {Altieri}, {Andreani}, {Aparicio-Villegas}, {Aussel}, {Ben{\'{\i}}tez},
  {Berta}, {Broadhurst}, {Cabrera-Ca{\~n}o}, {Castander}, {Cava},
  {Cervi{\~n}o}, {Chulani}, {Cimatti}, {Crist{\'o}bal-Hornillos}, {Daddi},
  {Dominguez}, {Elbaz}, {Fern{\'a}ndez-Soto}, {F{\"o}rster Schreiber},
  {Genzel}, {G{\'o}mez}, {Gonz{\'a}lez Delgado}, {Grazian}, {Gruppioni},
  {Herreros}, {Iglesias Groth}, {Infante}, {Lutz}, {Magnelli}, {Magdis},
  {Maiolino}, {M{\'a}rquez}, {Mart{\'{\i}}nez}, {Masegosa}, {Moles}, {Molino},
  {Nordon}, {Del Olmo}, {Perea}, {Poglitsch}, {Popesso}, {Pozzi}, {Prada},
  {Quintana}, {Riguccini}, {Rodighiero}, {Saintonge}, {S{\'a}nchez}, {Santini},
  {Shao}, {Sturm}, {Tacconi}, \& {Valtchanov}}]{Bongiovanni10}
{Bongiovanni}, A., et al. 2010, \aap, 519, L4+

\bibitem[{{Brooks} {et~al.}(2009){Brooks}, {Governato}, {Quinn}, {Brook}, \&
  {Wadsley}}]{Brooks09}
{Brooks}, A.~M., {Governato}, F., {Quinn}, T., {Brook}, C.~B., \& {Wadsley}, J.
  2009,  \apj, 694, 396

\bibitem[{{Bruzual} \& {Charlot}(2003)}]{Bruzual03}
{Bruzual}, G., \& {Charlot}, S. 2003, \mnras, 344, 1000

\bibitem[{{Cassata} {et~al.}(2011)}]{Cassata11}
{Cassata}, P., et al. 2011, \aap, 525, 143


\bibitem[{{Charlot} \& {Fall}(1993)}]{Charlot93}
{Charlot}, S., \& {Fall}, S.~M. 1993, 
  \apj, 415, 580

\bibitem[{{Ciardullo} {et~al.}(2011){Ciardullo}, {Gronwall}, {Wolf},
  {McCathran}, {Bond}, {Gawiser}, {Guaita}, {Feldmeier}, {Treister}, {Padilla},
  {Francke}, {Matkovic}, {Altmann}, \& {Herrera}}]{Ciardullo11}
{Ciardullo}, R., et al. 2011, ArXiv
  e-prints

\bibitem[{{Cowie} {et~al.}(2010){Cowie}, {Barger}, \& {Hu}}]{Cowie10}
{Cowie}, L.~L., {Barger}, A.~J., \& {Hu}, E.~M. 2010, 
  \apj, 711, 928

\bibitem[{{Cowie} {et~al.}(2011){Cowie}, {Barger}, \& {Hu}}]{Cowie11}
---. 2011,  \apj, 738, 136

\bibitem[{{Cowie} \& {Hu}(1998)}]{Cowie98}
{Cowie}, L.~L., \& {Hu}, E.~M. 1998,  \aj, 115, 1319

\bibitem[{{Cuby} {et~al.}(2007){Cuby}, {Hibon}, {Lidman}, {Le F{\`e}vre},
  {Gilmozzi}, {Moorwood}, \& {van der Werf}}]{Cuby07}
{Cuby}, J.-G., et al. 2007,  \aap, 461, 911

\bibitem[{{Dav{\'e}} {et~al.}(1999){Dav{\'e}}, {Hernquist}, {Katz}, \&
  {Weinberg}}]{Dave99}
{Dav{\'e}}, R., {Hernquist}, L., {Katz}, N., \& {Weinberg}, D.~H. 1999,  ApJ, 511, 521

\bibitem[{{Dawson} {et~al.}(2004){Dawson}, {Rhoads}, {Malhotra}, {Stern},
  {Dey}, {Spinrad}, {Jannuzi}, {Wang}, \& {Landes}}]{Dawson04}
{Dawson}, S., et al. 2004,
\apj, 617, 707

\bibitem[{{Dayal} \& {Libeskind}(2011)}]{Dayal11b}
{Dayal}, P., \& {Libeskind}, N.~I. 2012, \mnras, 419, L9

\bibitem[{{Dayal} {et~al.}(2008){Dayal}, {Ferrara}, \& {Gallerani}}]{Dayal08}
{Dayal}, P., {Ferrara}, A., \& {Gallerani}, S. 2008,  \mnras, 389, 1683

\bibitem[{{Dayal} {et~al.}(2011){Dayal}, {Maselli}, \& {Ferrara}}]{Dayal11}
{Dayal}, P., {Maselli}, A., \& {Ferrara}, A. 2011,  \mnras, 410, 830

\bibitem[{{Deharveng} {et~al.}(2008){Deharveng}, {Small}, {Barlow},
  {P{\'e}roux}, {Milliard}, {Friedman}, {Martin}, {Morrissey}, {Schiminovich},
  {Forster}, {Seibert}, {Wyder}, {Bianchi}, {Donas}, {Heckman}, {Lee},
  {Madore}, {Neff}, {Rich}, {Szalay}, {Welsh}, \& {Yi}}]{Deharveng08}
{Deharveng}, J.-M., et al. 2008,
  \apj, 680, 1072

\bibitem[{{Dekel} \& {Birnboim}(2006)}]{Dekel06}
{Dekel}, A., \& {Birnboim}, Y. 2006, \mnras, 368, 2

\bibitem[{{Dekel} {et~al.}(2009){Dekel}, {Birnboim}, {Engel}, {Freundlich},
  {Goerdt}, {Mumcuoglu}, {Neistein}, {Pichon}, {Teyssier}, \&
  {Zinger}}]{Dekel09}
{Dekel}, A., et al. 2009,  \nat, 457, 451

\bibitem[{{Di Matteo} {et~al.}(2008){Di Matteo}, {Colberg}, {Springel},
  {Hernquist}, \& {Sijacki}}]{DiMatteo08}
{Di Matteo}, T., {Colberg}, J., {Springel}, V., {Hernquist}, L., \& {Sijacki},
  D. 2008, \apj, 676, 33

\bibitem[{{Di Matteo} {et~al.}(2005){Di Matteo}, {Springel}, \&
  {Hernquist}}]{DiMatteo05}
{Di Matteo}, T., {Springel}, V., \& {Hernquist}, L. 2005,  \nat, 433, 604

\bibitem[{{Dijkstra} {et~al.}(2007){Dijkstra}, {Lidz}, \&
  {Wyithe}}]{Dijkstra07}
{Dijkstra}, M., {Lidz}, A., \& {Wyithe}, J.~S.~B. 2007,  \mnras, 377, 1175

\bibitem[{{Dijkstra} \& {Loeb}(2009)}]{Dijkstra09}
{Dijkstra}, M., \& {Loeb}, A. 2009,  \mnras, 400, 1109

\bibitem[{{Faucher-Gigu{\`e}re} {et~al.}(2009){Faucher-Gigu{\`e}re}, {Lidz},
  {Zaldarriaga}, \& {Hernquist}}]{Faucher09}
{Faucher-Gigu{\`e}re}, C., {Lidz}, A., {Zaldarriaga}, M., \& {Hernquist}, L.
  2009, \apj, 703, 1416

\bibitem[{{Finkelstein} {et~al.}(2009){Finkelstein}, {Cohen}, {Malhotra}, \&
  {Rhoads}}]{Finkelstein09}
{Finkelstein}, S.~L., {Cohen}, S.~H., {Malhotra}, S., \& {Rhoads}, J.~E. 2009,
   \apj, 700, 276

\bibitem[{{Finkelstein} {et~al.}(2011){Finkelstein}, {Hill}, {Gebhardt},
  {Adams}, {Blanc}, {Papovich}, {Ciardullo}, {Drory}, {Gawiser}, {Gronwall},
  {Schneider}, \& {Tran}}]{Finkelstein11}
{Finkelstein}, S.~L., et al. 2011, 
  \apj, 729, 140

\bibitem[{{Fujita} {et~al.}(2009)}]{Fujita09}
{Fujita}, A., {Martin}, C.~L., {Mac Low}, M.-M., {New}, K.~C.~B., \& {Weaver}, R.
2009, \apj, 698, 693
	
\bibitem[{{Fynbo} {et~al.}(2003){Fynbo}, {Ledoux}, {M{\"o}ller}, {Thomsen}, \&
  {Burud}}]{Fynbo03}
{Fynbo}, J.~P.~U., {Ledoux}, C., {M{\"o}ller}, P., {Thomsen}, B., \& {Burud},
  I. 2003,  \aap, 407, 147

\bibitem[{{Fynbo} {et~al.}(2001){Fynbo}, {M{\"o}ller}, \& {Thomsen}}]{Fynbo01}
{Fynbo}, J.~U., {M{\"o}ller}, P., \& {Thomsen}, B. 2001,  \aap, 374,
  443

\bibitem[{{Gawiser} {et~al.}(2007){Gawiser}, {Francke}, {Lai}, {Schawinski},
  {Gronwall}, {Ciardullo}, {Quadri}, {Orsi}, {Barrientos}, {Blanc}, {Fazio},
  {Feldmeier}, {Huang}, {Infante}, {Lira}, {Padilla}, {Taylor}, {Treister},
  {Urry}, {van Dokkum}, \& {Virani}}]{Gawiser07}
{Gawiser}, E., et al. 2007, \apj, 671, 278

\bibitem[{{Gawiser} {et~al.}(2006){Gawiser}, {van Dokkum}, {Gronwall},
  {Ciardullo}, {Blanc}, {Castander}, {Feldmeier}, {Francke}, {Franx},
  {Haberzettl}, {Herrera}, {Hickey}, {Infante}, {Lira}, {Maza}, {Quadri},
  {Richardson}, {Schawinski}, {Schirmer}, {Taylor}, {Treister}, {Urry}, \&
  {Virani}}]{Gawiser06}
{Gawiser}, E., et al. 2006, \apjl, 642, L13

\bibitem[{{Goerdt} {et~al.}(2010){Goerdt}, {Dekel}, {Sternberg}, {Ceverino},
  {Teyssier}, \& {Primack}}]{Goerdt10}
{Goerdt}, T., et al. 2010,  \mnras, 407, 613

\bibitem[{{Gronwall} {et~al.}(2007){Gronwall}, {Ciardullo}, {Hickey},
  {Gawiser}, {Feldmeier}, {van Dokkum}, {Urry}, {Herrera}, {Lehmer}, {Infante},
  {Orsi}, {Marchesini}, {Blanc}, {Francke}, {Lira}, \& {Treister}}]{Gronwall07}
{Gronwall}, C., et al. 2007,  \apj, 667, 79

\bibitem[{{Guaita} {et~al.}(2010){Guaita}, {Gawiser}, {Padilla}, {Francke},
  {Bond}, {Gronwall}, {Ciardullo}, {Feldmeier}, {Sinawa}, {Blanc}, \&
  {Virani}}]{Guaita10}
{Guaita}, L., et al. 2010,  \apj, 714, 255

\bibitem[{Haardt \& Madau (1996)}]{Haardt96}
Haardt, F., \& Madau, P. 1996, ApJ, 461, 20

\bibitem[{{Haiman} (2002)}]{Haiman02}
{Haiman}, Z. 2002, \apjl, 576, L1

\bibitem[{{Hayes} {et~al.}(2007){Hayes}, {{\"O}stlin}, {Atek}, {Kunth},
  {Mas-Hesse}, {Leitherer}, {Jim{\'e}nez-Bail{\'o}n}, \& {Adamo}}]{Hayes07}
{Hayes}, M., et al. 2007, \mnras, 382, 1465

\bibitem[{{Hayes} {et~al.}(2010){Hayes}, {{\"O}stlin}, {Schaerer}, {Mas-Hesse},
  {Leitherer}, {Atek}, {Kunth}, {Verhamme}, {de Barros}, \&
  {Melinder}}]{Hayes10}
{Hayes}, M., et al. 2010,  \nat, 464, 562

\bibitem[{{Hayes} {et~al.}(2011){Hayes}, {Schaerer}, {{\"O}stlin}, {Mas-Hesse},
  {Atek}, \& {Kunth}}]{Hayes11}
{Hayes}, M., et al. 2011,  \apj, 730, 8

\bibitem[{{Hernquist} \& {Katz}(1989)}]{Hernquist89}
{Hernquist}, L., \& {Katz}, N. 1989,  \apjs, 70, 419

\bibitem[{{Hockney} \& {Eastwood}(1981)}]{Hockney81}
{Hockney}, R.~W., \& {Eastwood}, J.~W. 1981, {Computer Simulation Using
  Particles}, ed. {Hockney, R.~W.~\& Eastwood, J.~W.}, {Computer Simulation
  Using Particles}

\bibitem[{{Hopkins} \& {Beacom}(2006)}]{Hopkins2006}
{Hopkins}, A.~M., \& {Beacom}, J.~F. 2006,  \apj, 651, 142

\bibitem[{{Hopkins} {et~al.}(2006){Hopkins}, {Hernquist}, {Cox}, {Di Matteo},
  {Robertson}, \& {Springel}}]{Hopkins06}
{Hopkins}, P.~F., et al. 2006,  \apjs, 163, 1

\bibitem[{{Horton} {et~al.}(2004){Horton}, {Parry}, {Bland-Hawthorn}, {Cianci},
  {King}, {McMahon}, \& {Medlen}}]{Horton04}
{Horton}, A., {Parry}, I., {Bland-Hawthorn}, J., {Cianci}, S., {King}, D.,
  {McMahon}, R., \& {Medlen}, S. 2004, in Society of Photo-Optical
  Instrumentation Engineers (SPIE) Conference Series, Vol. 5492, Society of
  Photo-Optical Instrumentation Engineers (SPIE) Conference Series, ed.
  {A.~F.~M.~Moorwood \& M.~Iye}, 1022--1032

\bibitem[{{Hoyle} \& {Lyttleton}(1941)}]{Hoyle41}
{Hoyle}, F., \& {Lyttleton}, R.~A. 1941,  \mnras, 101, 227

\bibitem[{{Hu} \& {Cowie}(2006)}]{Hu2006}
{Hu}, E.~M., \& {Cowie}, L.~L. 2006, \nat,
  440, 1145

\bibitem[{{Hu} {et~al.}(2010){Hu}, {Cowie}, {Barger}, {Capak}, {Kakazu}, \&
  {Trouille}}]{Hu2010}
{Hu}, E.~M., et al. 2010, 
  \apj, 725, 394

\bibitem[{{Hu} {et~al.}(2004){Hu}, {Cowie}, {Capak}, {McMahon}, {Hayashino}, \&
  {Komiyama}}]{Hu2004}
{Hu}, E.~M., et al. 2004, \aj, 127, 563

\bibitem[{{Hu} {et~al.}(1998){Hu}, {Cowie}, \& {McMahon}}]{Hu98}
{Hu}, E.~M., {Cowie}, L.~L., \& {McMahon}, R.~G. 1998, \apjl, 502, L99

\bibitem[{{Hu} {et~al.}(2002){Hu}, {Cowie}, {McMahon}, {Capak}, {Iwamuro},
  {Kneib}, {Maihara}, \& {Motohara}}]{Hu2002}
{Hu}, E.~M., et al. 2002, \apjl, 568, L75

\bibitem[{{Hu} \& {McMahon}(1996)}]{Hu96}
{Hu}, E.~M., \& {McMahon}, R.~G. 1996, \nat, 382, 231

\bibitem[{{Hui} \& {Gnedin}(1997)}]{Hui97}
{Hui}, L., \& {Gnedin}, N.~Y. 1997, \mnras, 292, 27

\bibitem[{{Iliev} {et~al.}(2008)}]{Iliev08}
{Iliev}, I.~T., {Shapiro}, P.~R., {McDonald}, P., {Mellema}, G., \& {Pen}, U.-L.
2008, \mnras, 391, 63

\bibitem[{{Iye} {et~al.}(2006){Iye}, {Ota}, {Kashikawa}, {Furusawa},
  {Hashimoto}, {Hattori}, {Matsuda}, {Morokuma}, {Ouchi}, \&
  {Shimasaku}}]{Iye06}
{Iye}, M., et al. 2006, \nat, 443, 186

\bibitem[{{Kashikawa} {et~al.}(2006){Kashikawa}, {Shimasaku}, {Malkan}, {Doi},
  {Matsuda}, {Ouchi}, {Taniguchi}, {Ly}, {Nagao}, {Iye}, {Motohara},
  {Murayama}, {Murozono}, {Nariai}, {Ohta}, {Okamura}, {Sasaki}, {Shioya}, \&
  {Umemura}}]{Kashikawa06}
{Kashikawa}, N., et al. 2006, \apj, 648, 7

\bibitem[{{Katz} {et~al.}(2003){Katz}, {Keres}, {Dave}, \& {Weinberg}}]{Katz03}
{Katz}, N., {Keres}, D., {Dave}, R., \& {Weinberg}, D.~H. 2003, in Astrophysics
  and Space Science Library, Vol. 281, The IGM/Galaxy Connection. The
  Distribution of Baryons at z=0, ed. {J.~L.~Rosenberg \& M.~E.~Putman}, 185--+

\bibitem[{{Katz} {et~al.}(1996){Katz}, {Weinberg}, {Hernquist}, \&
  {Miralda-Escude}}]{Katz96}
{Katz}, N., {Weinberg}, D.~H., {Hernquist}, L., \& {Miralda-Escude}, J. 1996,
  \apjl, 457, L57

\bibitem[{{Keel} {et~al.}(2009)}]{Keel09}
{Keel}, W.~C., {White}, III, R.~E., {Chapman}, S., \& {Windhorst}, R.~A. 2009, \apj, 138, 986
	
\bibitem[{{Kennicutt} (1998)}]{Kennicutt98}
{Kennicutt}, Jr., R.~C. 1998, \araa, 36, 189

\bibitem[{{Kere{\v s}} {et~al.}(2009){Kere{\v s}}, {Katz}, {Fardal},
  {Dav{\'e}}, \& {Weinberg}}]{Keres09}
{Kere{\v s}}, D., {Katz}, N., {Fardal}, M., {Dav{\'e}}, R., \& {Weinberg},
  D.~H. 2009,  \mnras, 395, 160

\bibitem[{{Kere{\v s}} {et~al.}(2005){Kere{\v s}}, {Katz}, {Weinberg}, \&
  {Dav{\'e}}}]{Keres05}
{Kere{\v s}}, D., {Katz}, N., {Weinberg}, D.~H., \& {Dav{\'e}}, R. 2005,  \mnras, 363, 2

\bibitem[{{Kodaira} {et~al.}(2003){Kodaira}, {Taniguchi}, {Kashikawa}, {Kaifu},
  {Ando}, {Karoji}, {Ajiki}, {Akiyama}, {Aoki}, {Doi}, {Fujita}, {Furusawa},
  {Hayashino}, {Imanishi}, {Iwamuro}, {Iye}, {Kawabata}, {Kobayashi}, {Kodama},
  {Komiyama}, {Kosugi}, {Matsuda}, {Miyazaki}, {Mizumoto}, {Motohara},
  {Murayama}, {Nagao}, {Nariai}, {Ohta}, {Ohyama}, {Okamura}, {Ouchi},
  {Sasaki}, {Sekiguchi}, {Shimasaku}, {Shioya}, {Takata}, {Tamura}, {Terada},
  {Umemura}, {Usuda}, {Yagi}, {Yamada}, {Yasuda}, \& {Yoshida}}]{Kodaira03}
{Kodaira}, K., et al. 2003,
  \pasj, 55, L17

\bibitem[{{Komatsu} {et~al.}(2009){Komatsu}, {Dunkley}, {Nolta}, {Bennett},
  {Gold}, {Hinshaw}, {Jarosik}, {Larson}, {Limon}, {Page}, {Spergel},
  {Halpern}, {Hill}, {Kogut}, {Meyer}, {Tucker}, {Weiland}, {Wollack}, \&
  {Wright}}]{Komatsu09}
{Komatsu}, E., et al. 2009,
 \apjs, 180, 330

\bibitem[{{Lai} {et~al.}(2007){Lai}, {Huang}, {Fazio}, {Cowie}, {Hu}, \&
  {Kakazu}}]{Lai07}
{Lai}, K., et al. 2007, \apj, 655, 704

\bibitem[{{Lai} {et~al.}(2008){Lai}, {Huang}, {Fazio}, {Gawiser}, {Ciardullo},
  {Damen}, {Franx}, {Gronwall}, {Labbe}, {Magdis}, \& {van Dokkum}}]{Lai08}
{Lai}, K., et al. 2008,  \apj, 674, 70

\bibitem[{{Laursen} {et~al.}(2009){Laursen}, {Sommer-Larsen}, \&
  {Andersen}}]{Laursen09b}
{Laursen}, P., {Sommer-Larsen}, J., \& {Andersen}, A.~C. 2009,  \apj, 704, 1640

\bibitem[{{Laursen} {et~al.}(2011){Laursen}, {Sommer-Larsen}, \&
  {Razoumov}}]{Laursen11}
{Laursen}, P., {Sommer-Larsen}, J., \& {Razoumov}, A.~O. 2011, \apj, 728, 52

\bibitem[{{Lehnert} {et~al.}(2010){Lehnert}, {Nesvadba}, {Cuby}, {Swinbank},
  {Morris}, {Cl{\'e}ment}, {Evans}, {Bremer}, \& {Basa}}]{Lehnert10}
{Lehnert}, M.~D., et al. 2010,  \nat,
  467, 940

\bibitem[{{Li} {et~al.}(2007){Li}, {Hernquist}, {Robertson}, {Cox}, {Hopkins},
  {Springel}, {Gao}, {Di Matteo}, {Zentner}, {Jenkins}, \& {Yoshida}}]{Li07}
{Li}, Y., et al. 2007, \apj, 665, 187

\bibitem[{{Li} {et~al.}(2008){Li}, {Hopkins}, {Hernquist}, {Finkbeiner}, {Cox},
  {Springel}, {Jiang}, {Fan}, \& {Yoshida}}]{Li08}
{Li}, Y., et al. 2008,  \apj, 678, 41

\bibitem[{{Maier} {et~al.}(2003){Maier}, {Meisenheimer}, {Thommes},
  {Hippelein}, {R{\"o}ser}, {Fried}, {von Kuhlmann}, {Phleps}, \&
  {Wolf}}]{Maier03}
{Maier}, C., et al. 2003,
  \aap, 402, 79

\bibitem[{{Malhotra} \& {Rhoads}(2004)}]{Malhotra04}
{Malhotra}, S., \& {Rhoads}, J.~E. 2004, \apjl, 617, L5

\bibitem[{{Malhotra} {et~al.}(2011){Malhotra}, {Rhoads}, {Finkelstein},
  {Hathi}, {Nilsson}, {McLinden}, \& {Pirzkal}}]{Malhotra11}
{Malhotra}, S., et al. 2011, arXiv:1106.2816

\bibitem[{{Mas-Hesse} {et~al}(2003)}]{Mas-Hesse03}
{Mas-Hesse}, J.~M., {Kunth}, D., {Tenorio-Tagle}, G.,
	{Leitherer}, C., {Terlevich}, R.~J., \& {Terlevich}, E.
2003, \apj, 598, 858

\bibitem[{{Matsuda} {et~al.}(2011)}]{Matsuda11}
{Matsuda}, Y., et al. 2011, \mnras, 410, L13

\bibitem[{Mesinger} \& {Furlanetto}(2008)]{Mesinger08}
{Mesinger}, A., \& {Furlanetto}, S.~R.
2008, \mnras, 386, 1990

\bibitem[{McQuinn} {et~al.}(2007)]{McQuinn07}
{McQuinn}, M., {Hernquist}, L., {Zaldarriaga}, M., \& {Dutta}, S.
2007, \mnras, 381, 75

\bibitem[{{Nilsson} \& {M{\o}ller}(2011)}]{Nilsson11}
{Nilsson}, K.~K., \& {M{\o}ller}, P. 2011, \aap, 527, L7

\bibitem[{{Nilsson} {et~al.}(2007){Nilsson}, {M{\o}ller}, {M{\"o}ller},
  {Fynbo}, {Micha{\l}owski}, {Watson}, {Ledoux}, {Rosati}, {Pedersen}, \&
  {Grove}}]{Nilsson07}
{Nilsson}, K.~K., et al. 2007,  \aap, 471, 71

\bibitem[{{Nilsson} {et~al.}(2011){Nilsson}, {{\"O}stlin}, {M{\o}ller},
  {M{\"o}ller-Nilsson}, {Tapken}, {Freudling}, \& {Fynbo}}]{Nilsson11b}
{Nilsson}, K.~K., et al. 2011, \aap, 529, A9

\bibitem[{{Nilsson} {et~al.}(2009){Nilsson}, {Tapken}, {M{\o}ller},
  {Freudling}, {Fynbo}, {Meisenheimer}, {Laursen}, \& {{\"O}stlin}}]{Nilsson09}
{Nilsson}, K.~K., et al. 2009,
 \aap, 498, 13

\bibitem[{{Ocvirk} {et~al.}(2008){Ocvirk}, {Pichon}, \& {Teyssier}}]{Ocvirk08}
{Ocvirk}, P., {Pichon}, C., \& {Teyssier}, R. 2008, \mnras, 390, 1326

\bibitem[{{Ono} {et~al.}(2010{\natexlab{a}}){Ono}, {Ouchi}, {Shimasaku},
  {Akiyama}, {Dunlop}, {Farrah}, {Lee}, {McLure}, {Okamura}, \&
  {Yoshida}}]{Ono10A}
{Ono}, Y., et al.
  2010{\natexlab{a}},
  \mnras, 402, 1580

\bibitem[{{Ono} {et~al.}(2010{\natexlab{b}}){Ono}, {Ouchi}, {Shimasaku},
  {Dunlop}, {Farrah}, {McLure}, \& {Okamura}}]{Ono10B}
{Ono}, Y., et al. 2010{\natexlab{b}}, \apj, 724, 1524

\bibitem[{{Ono} {et~al.}(2012)}]{Ono12}
{Ono}, Y., et al. 2012, \apj, 744, 83

\bibitem[{{Osterbrock} \& {Ferland}(2006)}]{Osterbrock06}
{Osterbrock}, D.~E., \& {Ferland}, G.~J. 2006, {Astrophysics of gaseous nebulae
  and active galactic nuclei}, ed. {Osterbrock, D.~E.,~\& Ferland, G.~J.},
  {Astrophysics of gaseous nebulae and active galactic nuclei}

\bibitem[{{{\"O}stlin} {et~al.}(2009){{\"O}stlin}, {Hayes}, {Kunth},
  {Mas-Hesse}, {Leitherer}, {Petrosian}, \& {Atek}}]{Ostlin09}
{{\"O}stlin}, G., et al. 2009,  \aj, 138, 923

\bibitem[{{Ota} {et~al.}(2008){Ota}, {Iye}, {Kashikawa}, {Shimasaku},
  {Kobayashi}, {Totani}, {Nagashima}, {Morokuma}, {Furusawa}, {Hattori},
  {Matsuda}, {Hashimoto}, \& {Ouchi}}]{Ota08}
{Ota}, K., et al. 2008,  \apj, 677, 12

\bibitem[{{Ouchi} {et~al.}(2008){Ouchi}, {Shimasaku}, {Akiyama}, {Simpson},
  {Saito}, {Ueda}, {Furusawa}, {Sekiguchi}, {Yamada}, {Kodama}, {Kashikawa},
  {Okamura}, {Iye}, {Takata}, {Yoshida}, \& {Yoshida}}]{Ouchi08}
{Ouchi}, M., et al. 2008,  \apjs, 176, 301

\bibitem[{{Ouchi} {et~al.}(2003){Ouchi}, {Shimasaku}, {Furusawa}, {Miyazaki},
  {Doi}, {Hamabe}, {Hayashino}, {Kimura}, {Kodaira}, {Komiyama}, {Matsuda},
  {Miyazaki}, {Nakata}, {Okamura}, {Sekiguchi}, {Shioya}, {Tamura},
  {Taniguchi}, {Yagi}, \& {Yasuda}}]{Ouchi03}
{Ouchi}, M., et al. 2003, \apj, 582, 60

\bibitem[{{Ouchi} {et~al.}(2010){Ouchi}, {Shimasaku}, {Furusawa}, {Saito},
  {Yoshida}, {Akiyama}, {Ono}, {Yamada}, {Ota}, {Kashikawa}, {Iye}, {Kodama},
  {Okamura}, {Simpson}, \& {Yoshida}}]{Ouchi10}
{Ouchi}, M., et al. 2010,
 \apj, 723, 869

\bibitem[{{Partridge} \& {Peebles}(1967)}]{Partridge67}
{Partridge}, R.~B. \& {Peebles}, P.~J.~E. 1967, 
  \apj, 147, 868

\bibitem[{{Pentericci} {et~al.}(2009){Pentericci}, {Grazian}, {Fontana},
  {Castellano}, {Giallongo}, {Salimbeni}, \& {Santini}}]{Pentericci09}
{Pentericci}, L., et al. 2009, \aap, 494, 553

\bibitem[{Pentericci} {et~al.}(2011)]{Pentericci11}
Pentericci, L., et al. 2011, \apj, 743, 132

\bibitem[{{Pirzkal} {et~al.}(2007){Pirzkal}, {Malhotra}, {Rhoads}, \&
  {Xu}}]{Pirzkal07}
{Pirzkal}, N., {Malhotra}, S., {Rhoads}, J.~E., \& {Xu}, C. 2007,
  \apj, 667, 49

\bibitem[{Raiter} {et~al.}(2010)]{Raiter10}
{Raiter}, A., {Schaerer}, D., \& {Fosbury}, R.~A.~E.
2010, \aap, 523, 64

\bibitem[{{Rhoads} {et~al.}(2003){Rhoads}, {Dey}, {Malhotra}, {Stern},
  {Spinrad}, {Jannuzi}, {Dawson}, {Brown}, \& {Landes}}]{Rhoads03}
{Rhoads}, J.~E., et al. 2003,
  \aj, 125, 1006

\bibitem[{{Rhoads} {et~al.}(2000){Rhoads}, {Malhotra}, {Dey}, {Stern},
  {Spinrad}, \& {Jannuzi}}]{Rhoads00}
{Rhoads}, J.~E., et al. 2000, \apjl, 545, L85

\bibitem[{Salpeter(1955)}]{Salpeter55}
Salpeter, E.~E. 1955, ApJ, 121, 161

\bibitem[{{Salvadori} {et~al.}(2010){Salvadori}, {Dayal}, \&
  {Ferrara}}]{Salvadori10}
{Salvadori}, S., {Dayal}, P., \& {Ferrara}, A. 2010, \mnras, 407, L1

\bibitem[{{Santos}(2004)}]{Santos04}
{Santos}, M.~R. 2004,  \mnras, 349, 1137

\bibitem[{{Schaerer}(2003)}]{Schaerer03}
{Schaerer}, D. 2003, \aap, 397, 527

\bibitem[{{Schenker}{et~al.}(2012)}]{Schenker12}
{Schenker}, M.~A., et al. 2012, \apj, 744, 179

\bibitem[{{Schmidt}(1959)}]{Schmidt59}
{Schmidt}, M. 1959, \apj, 129, 243

\bibitem[{{Sheth} \& {Tormen} (1999)}]{Sheth99}
{Sheth}, R.~K., \& {Tormen}, G.
1999, \mnras, 308, 119 

\bibitem[{{Shimasaku} {et~al.}(2006){Shimasaku}, {Kashikawa}, {Doi}, {Ly},
  {Malkan}, {Matsuda}, {Ouchi}, {Hayashino}, {Iye}, {Motohara}, {Murayama},
  {Nagao}, {Ohta}, {Okamura}, {Sasaki}, {Shioya}, \& {Taniguchi}}]{Shimasaku06}
{Shimasaku}, K., et al. 2006,  \pasj, 58, 313

\bibitem[{{Springel}(2005)}]{Springel05e}
{Springel}, V. 2005, MNRAS, 364,
  1105

\bibitem[{{Springel} {et~al.}(2005{\natexlab{a}}){Springel}, {Di Matteo}, \&
  {Hernquist}}]{Springel05a}
{Springel}, V., {Di Matteo}, T., \& {Hernquist}, L. 2005{\natexlab{a}}, ApJL,
  620, L79

\bibitem[{{Springel} {et~al.}(2005{\natexlab{b}}){Springel}, {Di Matteo}, \&
  {Hernquist}}]{Springel05d}
---. 2005{\natexlab{b}}, MNRAS, 361, 776

\bibitem[{{Springel} \& {Hernquist}(2002)}]{Springel02}
{Springel}, V., \& {Hernquist}, L. 2002, MNRAS, 333, 649

\bibitem[{{Springel} \& {Hernquist}(2003)}]{Springel03a}
---. 2003, MNRAS, 339, 289

\bibitem[{{Springel} {et~al.}(2008){Springel}, {Wang}, {Vogelsberger},
  {Ludlow}, {Jenkins}, {Helmi}, {Navarro}, {Frenk}, \& {White}}]{Springel08a}
{Springel}, V., et al. 2008,
 MNRAS, 391, 1685

\bibitem[{{Springel} {et~al.}(2001){Springel}, {Yoshida}, \&
  {White}}]{Springel01}
{Springel}, V., {Yoshida}, N., \& {White}, S.~D.~M. 2001, New Astronomy, 6,
  79

\bibitem[{{Stark} {et~al.}(2007){Stark}, {Ellis}, {Richard}, {Kneib}, {Smith},
  \& {Santos}}]{Stark07}
{Stark}, D.~P., et al. 2007, \apj,
  663, 10

\bibitem[{{Stark} {et~al.}(2010)}]{Stark10}
{Stark}, D.~P., {Ellis}, R.~S., {Chiu}, K., {Ouchi}, M. \& {Bunker}, A.
2010, \mnras, 408, 1628

\bibitem[{{Stark} {et~al.}(2011)}]{Stark11}
{Stark}, D.~P., {Ellis}, R.~S., \& {Ouchi}, M.
2011, \apjl, 728, L2

\bibitem[{{Steidel} {et~al.}(2000){Steidel}, {Adelberger}, {Shapley},
  {Pettini}, {Dickinson}, \& {Giavalisco}}]{Steidel00}
{Steidel}, C.~C., et al. 2000,  \apj, 532, 170

\bibitem[{{Stern} {et~al.}(2005){Stern}, {Yost}, {Eckart}, {Harrison},
  {Helfand}, {Djorgovski}, {Malhotra}, \& {Rhoads}}]{Stern05}
{Stern}, D., et al. 2005, \apj, 619,
  12

\bibitem[{{Taniguchi} {et~al.}(2005){Taniguchi}, {Ajiki}, {Nagao}, {Shioya},
  {Murayama}, {Kashikawa}, {Kodaira}, {Kaifu}, {Ando}, {Karoji}, {Akiyama},
  {Aoki}, {Doi}, {Fujita}, {Furusawa}, {Hayashino}, {Iwamuro}, {Iye},
  {Kobayashi}, {Kodama}, {Komiyama}, {Matsuda}, {Miyazaki}, {Mizumoto},
  {Morokuma}, {Motohara}, {Nariai}, {Ohta}, {Ohyama}, {Okamura}, {Ouchi},
  {Sasaki}, {Sato}, {Sekiguchi}, {Shimasaku}, {Tamura}, {Umemura}, {Yamada},
  {Yasuda}, \& {Yoshida}}]{Taniguchi05}
{Taniguchi}, Y., et al.
  2005, \pasj, 57, 165

\bibitem[{{Vanzella} {et~al.}(2011)}]{Vanzella11}
{Vanzella}, E., et al. 2011, \apjl, 730, L35

\bibitem[{{Wadepuhl} \& {Springel}(2011)}]{Wadepuhl11}
{Wadepuhl}, M., \& {Springel}, V. 2011, \mnras, 410, 1975

\bibitem[{{Willis} {et~al.}(2008){Willis}, {Courbin}, {Kneib}, \&
  {Minniti}}]{Willis08}
{Willis}, J.~P., {Courbin}, F., {Kneib}, J.-P., \& {Minniti}, D. 2008, \mnras, 384, 1039

\bibitem[{{Xu} (1995)}]{Xu95}
{Xu}, G. 1995,  \apjs, 98, 355

\bibitem[{{Yajima} {et~al.}(2011){Yajima}, {Li}, {Zhu}, \&
  {Abel}}]{Yajima11A}
{Yajima}, H., {Li}, Y., {Zhu}, Q., \& {Abel}, T. 2011, arXiv: 1109.4891



\bibitem[{{Yamada} {et~al.}(2012)}]{Yamada12}
{Yamada}, T., et al. 2012, arXiv: 1203.3633

\bibitem[{{Zheng} {et~al.}(2010){Zheng}, {Cen}, {Trac}, \&
  {Miralda-Escud{\'e}}}]{Zheng10}
{Zheng}, Z., {Cen}, R., {Trac}, H., \& {Miralda-Escud{\'e}}, J. 2010,
  \apj, 716, 574

\bibitem[{{Zheng} \& {Miralda-Escud{\'e}}(2002)}]{Zheng02}
{Zheng}, Z. \& {Miralda-Escud{\'e}}, J. 2002,  \apj, 578, 33


\end{thebibliography}

\end{document}